\pdfoutput=1
\documentclass[aps,prd,twocolumn,nofootinbib,preprintnumbers,superscriptaddress,10pt]{revtex4-2}

\usepackage[pdfborder={0 0 0},bookmarks=true]{hyperref}
\usepackage{graphicx,color,xcolor}
\definecolor{darkblue}{rgb}{0,0,.5}
\definecolor{darkgreen}{rgb}{0,0.5,.5}
\definecolor{darkyellow}{rgb}{0.5,0.5,0}
\definecolor{fhl}{rgb}{1,0,0}
\hypersetup{colorlinks=true, breaklinks=true, linkcolor=darkblue, menucolor=darkblue, urlcolor=darkblue, linktocpage=true}
\usepackage[english]{babel}
\usepackage{amssymb,amsfonts,amsmath}
\usepackage{bm}
\usepackage{bbm}
\usepackage{mathtools}
\usepackage{tikz}
\usetikzlibrary{arrows,shapes,backgrounds,calc,positioning,tikzmark,patterns,automata,decorations.markings}
\usepackage{verbatim}
\usepackage{picture}
\usepackage{cancel}
\usepackage{tabularx}
\usepackage{tensor}
\usepackage{fancyhdr}
\usepackage{sidecap}
\usepackage{scalerel}
\usepackage{relsize}
\usepackage[absolute,overlay]{textpos}
\usepackage[customcolors]{hf-tikz}
\usepackage{cals}
\usepackage{enumitem}
\setlist{nolistsep}
\usepackage{pifont}
\usepackage[export]{adjustbox} 
\usepackage[most]{tcolorbox}
\usepackage{varwidth}
\usepackage{xfrac}
\tcbuselibrary{skins}

\newcommand{\ie}{{\emph{i.e.}}}
\newcommand{\eg}{{\emph{e.g.}}}

\newcommand{\be}{\begin{equation}}
	\newcommand{\ee}{\end{equation}}
\newcommand{\ba}{\begin{eqnarray}}
	\newcommand{\ea}{\end{eqnarray}}
\newcommand{\baa}{\begin{array}}
	\newcommand{\eaa}{\end{array}}

\newcommand{\tr}{{\text{tr}}}
\newcommand{\SEE}{S_{\text{EE}}}
\newcommand{\dSEE}{\partial_\ell\SEE}
\newcommand{\dHn}[1]{\partial_\ell H_{#1}}

\newcommand{\vspa}{L L_\perp^{d-2}}

\newcommand{\tra}{\text{tr}}

\newcommand{\ack}[1]{{\color{red}{{\bf #1}}}}

\let\originalleft\left
\let\originalright\right
\renewcommand{\left}{\mathopen{}\mathclose\bgroup\originalleft}
\renewcommand{\right}{\aftergroup\egroup\originalright}

\newcommand{\e}{\operatorname{e}}

\newcommand{\SU}[1]{\operatorname{SU}(#1)}

\newcommand{\On}[1]{\operatorname{O}\left(#1\right)}

\newcommand{\of}[1]{\left(#1\right)}

\newcommand{\bof}[1]{\biggl(\bigg.#1\bigg.\biggr)}
\newcommand{\sbof}[1]{\Bigl(\Big.#1\Big.\Bigr)}
\newcommand{\sof}[1]{\bigl(\big.#1\big.\bigr)}
\newcommand{\ssof}[1]{(#1)}
\newcommand{\fof}[1]{\left[#1\right]}

\newcommand{\cof}[1]{\left\{#1\right\}}

\newcommand{\bcof}[1]{\biggl\{\bigg.#1\bigg.\biggr\}}

\newcommand{\sscof}[1]{\{#1\}}

\newcommand{\avof}[1]{\left\langle #1\right\rangle}

\newcommand{\ssavof}[1]{\small\langle #1\small\rangle}

\newcommand{\trace}{\operatorname{tr}}

\newcommand{\ii}{\mathrm{i}}
\newcommand{\idd}[2]{\mathrm{d}^{#2}#1}
\newcommand{\dd}{\mathrm{d}}
\newcommand{\DD}[1]{\mathcal{D}\bigl[#1\bigr]}

\newcommand{\partd}[2]{\frac{\partial #1}{\partial #2}}
\newcommand{\partdf}[3]{\left.\frac{\partial #1}{\partial #2}\right\vert_{#3}}

\newcommand{\spartd}[3]{\frac{\partial^{2} #1}{\partial #2\,\partial #3}}

\newcommand{\sorder}[1]{\mathcal{O}\big(#1\big)}

\newcommand{\abs}[1]{\left| #1\right|}

\newcommand{\sabs}[1]{\big| #1\big|}

\renewcommand*\hat[1]{\widehat{#1}}
\let\oldstackrel\stackrel
\renewcommand*\stackrel[2]{{\scriptstyle\oldstackrel{#1}{#2}}}
\definecolor{darkgreen}{RGB}{18,103,74}
\definecolor{emphcol}{RGB}{0,0,0}
\let\oldemph\emph
\renewcommand*\emph[1]{\oldemph{\textcolor{emphcol}{#1}}}
\let\oldstackrel\stackrel
\renewcommand*\stackrel[2]{{\scriptstyle\oldstackrel{#1}{#2}}}

\newcommand{\ucases}[1]{\begin{cases}#1\end{cases}}

\newcommand*\getscale[1]{%
  \begingroup
    \pgfgettransformentries{\scaleA}{\scaleB}{\scaleC}{\scaleD}{\whatevs}{\whatevs}%
    \pgfmathsetmacro{#1}{sqrt(abs(\scaleA*\scaleD-\scaleB*\scaleC))}%
    \expandafter
  \endgroup
  \expandafter\edef\expandafter#1\expandafter{#1}%
}

\newcommand{\vlatt}{N_t V}
\newcommand{\vrlatt}{2 N_t V}
\newcommand{\nrep}{\tilde{n}}

\newcolumntype{C}[1]{>{\centering\let\newline\\\arraybackslash\hspace{0pt}}m{#1}}

\makeatletter
\renewcommand{\p@subsection}{}
\renewcommand{\p@subsubsection}{}
\renewcommand{\p@paragraph}{}
\makeatother

\begin{document}

\title{Determination of thermodynamics from entanglement entropy \\ in the finite-density O(N) model}

\author{Niko Jokela}
\email{niko.jokela@helsinki.fi}
\affiliation{Department of Physics and Helsinki Institute of Physics,\\
P.O.~Box 64, FI-00014 University of Helsinki, Finland}

\author{Aatu Rajala}
\email{aatu.rajala@helsinki.fi}
\affiliation{Department of Physics and Helsinki Institute of Physics,\\
P.O.~Box 64, FI-00014 University of Helsinki, Finland}

\author{Tobias Rindlisbacher}
\email{tobias.rindlisbacher@helsinki.fi}
\affiliation{Department of Physics and Helsinki Institute of Physics,\\
P.O.~Box 64, FI-00014 University of Helsinki, Finland}

\begin{abstract}
We nonperturbatively compute R\'enyi entropies for strip-shaped subregions in the three-dimensional $\On{4}$ model at finite density on the lattice. By using a dual variable representation and a tailored worm algorithm, we circumvent the sign problem when sampling the grand canonical ensemble. In the limit of large subregions, we also establish a direct, quantitative relationship between the derivative of entanglement entropy with respect to the size of the entangling region and the thermal entropy density for general quantum field theories, providing a new way to study their thermodynamics. We corroborate this argument with our lattice results by demonstrating that, in the appropriate limit, the derivative of entanglement entropy satisfies the same Maxwell relation as the thermal entropy density. 
\end{abstract}

\preprint{HIP-2026-9/TH}

\pacs{xyz}

\maketitle

\section{Introduction}\label{sec:introduction}

In quantum field theory (QFT), entanglement arises naturally from conservation laws: each of them gives rise to a quantum number by which excitations of the QFT can be labeled and which represents the amount of conserved charge of the type that excitation carries. In interactions, these quantum numbers can only change in ways that respect the conservation laws. This means in particular that, whenever particles are created in pairs, their quantum numbers must be correlated in order to respect the conservation laws, \ie, if they carry non-trivial quantum numbers, they must form entangled pairs. This sort of correlation is a universal feature of interacting QFTs, and entanglement measures, such as entanglement entropy (EE), R\'enyi entropies, and asymmetries, provide a precise way to quantify these correlations and probe the properties of the QFT from different viewpoints. In particular, EE has emerged as a diagnostic tool in QFTs over the past two decades, capturing information about critical phenomena~\cite{Vidal:2002rm,Calabrese:2004eu}, renormalization group flows~\cite{Casini:2012ei,Nishioka:2018khk}, and thermodynamic properties~\cite{Swingle:2011np,Nakagawa:2010kjk,Jokela:2023rba,Bianchi:2026bxu}. Thanks to holographic~\cite{Nishioka:2018khk}, tensor-network~\cite{Coser:2013qda,Yang:2015rra,Bazavov:2017hzi,Cataldi:2023xki,Hayazaki:2025srr}, quantum Monte Carlo (MC)~\cite{Wang:2024edi,liu2026entanglementrenyinegativityfinitetemperature,Zhu:2026cls}, and lattice methods~\cite{Buividovich:2008kq,Nakagawa:2009jk,Nakagawa:2010kjk,Itou:2015cyu,Alba:2016bcp,Rabenstein:2018bri,Jokela:2023rba,Bulgarelli:2023ofi,Amorosso:2024leg,Amorosso:2024glf,Bulgarelli:2024onj,Bulgarelli:2024yrz,Bulgarelli:2025ewp,Amorosso:2026mdo, Amorosso:2026zkj}, it has become possible to evaluate EE and other entanglement measures in increasingly complex interacting theories.

Despite their conceptual importance, nonperturbative evaluations of EE at finite density remain exceptionally rare. Most existing results are confined to free fermions characterized by a Fermi surface~\cite{Wolf:2006zzb,Swingle:2009bf}, holographic models in the large-$N$ limit~\cite{Hartnoll:2012ux,Belin:2013uta,DiNunno:2021eyf}, or lower-dimensional spin chains~\cite{Korepin:2004zz,Laflorencie:2005duh}. Free bosonic theories are largely omitted in this context. At zero temperature, introducing a chemical potential in a noninteracting bosonic system leads to an ideal Bose--Einstein condensate. Because the particles in a condensate are distributed over space in a coherent way, the system is naturally in a highly entangled state~\cite{Simon:2001koo,Amico:2007ag}. Specifically, any two spatial regions within a finite-particle condensate are entangled, and this entanglement depends entirely on the macroscopic coherence rather than the physical distance between the regions. However, this spatial entanglement is constrained by the overall particle number conservation. The reduced density matrix for a spatial bipartition follows a simple binomial distribution, meaning that the EE scales only logarithmically with the total particle number~\cite{Ding:2009epm}. Consequently, in the thermodynamic limit, the extensive entropy density strictly vanishes, yielding no bulk thermodynamic information. While finite temperature does generate extensive thermal entropy in such free systems, capturing a rich, nonperturbative thermodynamic equation of state requires the introduction of interactions~\cite{Andersen:2003qj,Dupuis:2020fhh}. Therefore, interacting QFTs are needed for interesting bosonic entanglement thermodynamics.

Lattice field theory is one of the most widely used and successful approaches to nonperturbative studies of QFTs \cite{DeGrand,Gattringer:2010zz,Rothe:1992nt,Smit}. However, introducing a chemical potential to lattice formulations of interacting QFTs typically results in a complex action, leading to a severe sign problem that paralyzes traditional MC methods~\cite{deForcrand:2009zkb,Gattringer:2016kco,Aarts:2026uiu}. To overcome this issue, we turn to the $\On{N}$ model which can be re-expressed on the lattice in terms of integer-valued dual variables~\cite{Endres:2006xu,Gattringer:2012df,Gattringer:2012ap,Bruckmann:2015sua,Rindlisbacher:2015xku,Rindlisbacher:2016zht,Katz:2016azl,Rindlisbacher:2017ysn} for which the sign problem is absent. This dual representation can be efficiently simulated with a worm algorithm~\cite{Rindlisbacher:2015xku,Rindlisbacher:2016zht,Rindlisbacher:2017ysn}, granting us direct nonperturbative access to the finite-density regime. The $\On{N}$ model also provides a fertile testing ground for new ideas: despite its apparent simplicity, it exhibits rich infrared physics, including spontaneous symmetry breaking and the emergence of Goldstone modes. While our algorithm allows for general $N$ and arbitrary lattice dimensions $d\geq2$, in this first study, we focus on the case of $N=4$ in $d=2+1$ dimensions.

In this paper, which serves as the detailed technical companion to~\cite{Jokela:2026hwz}, we perform a lattice study of R\'enyi entropies, which can be used as estimators of EE, in the $\On{4}$ model at finite density. Using the replica trick~\cite{Calabrese:2004eu,Calabrese:2009qy}, we compute the derivative of EE with respect to the subregion width for strip-shaped subregions. To overcome the severe overlap problem present in lattice EE calculations, we adapt the entangling region boundary deformation method from~\cite{Jokela:2023rba} to our dual variable formalism of the $\On{N}$ model.
In addition to demonstrating these lattice techniques, we establish that EE can transcend its traditional role as an information-theoretic diagnostic tool by providing access to fundamental bulk thermodynamics. We show that the derivative of EE with respect to the subregion size precisely isolates the thermal entropy density in the limit of large subregions, \ie, if the linear extent of the subregions is larger than all relevant correlation lengths. With this, we are able to verify a generalized Maxwell relation and ultimately provide a novel, nonperturbative framework which could potentially enable us to extract the full equation of state directly from entanglement data. The method could, in principle, be extended to other interacting systems at both vanishing and finite temperatures and densities. 

The remainder of this paper is organized as follows: In section~\ref{sec:EEandreplica}, we review the formulation of the replica trick and its application to EE. Section~\ref{sect:EEthermo}  details the theoretical connection between entanglement thermodynamics and the Landau free energy. In section~\ref{sec:latticesetup}, we define the $\On{N}$ model at finite density on the lattice, introducing the dual variable representation and the worm algorithm. Section~\ref{sect:latt_EE} describes the specific numerical update algorithms used to extract EE from lattice simulations of the $\On{N}$ model in the dual variable formalism. In section~\ref{sec:select}, we detail our simulation parameters, and in section~\ref{sec:results}, we present our numerical results and physical conclusions. We conclude with the outlook in section~\ref{sec:outlook}, while transition probabilities and crosschecks of different update schemes are relegated to various appendices.

\section{Entanglement Entropy and the replica trick}\label{sec:EEandreplica}

\begin{figure}[!htb]
    \centering
    \includegraphics[width=0.4875\textwidth]{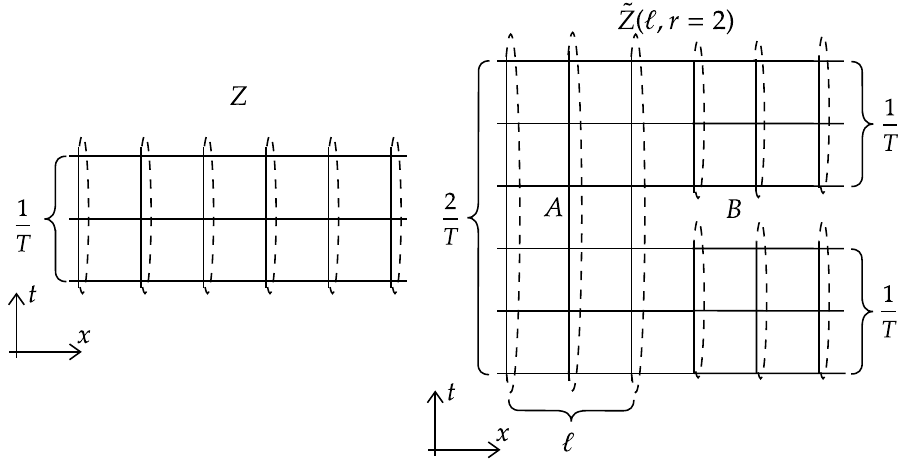}
    \caption{Sketches demonstrating the temporal topologies of the systems described by the two partition functions $Z$ and $\tilde{Z}\of{\ell,r}$ used in the replica method \eqref{eq:replicatrick}. On the left, we have the ordinary finite temperature partition function $Z$ at some temperature $T$, so that field configurations are $1/T$ periodic in the Euclidean time direction. On the right is the system described by $\tilde{Z}(\ell,r)$, consisting of two regions, $A$ and $B$, where region $A$ is a slab of width $\ell$ along the $x$ axis, and $B$ is the complement of $A$. In $A$ the fields are $r/T$ periodic, while in $B$ the total Euclidean time domain $r/T$ splits into $r$ disjoint parts, in each of which the fields are $1/T$ periodic. In the sketch we used $r=2$ to provide a concrete example and since $\tilde{Z}\of{\ell,r=2}$ is the partition function that we will end up simulating on the lattice. The sketches illustrate the systems corresponding to $Z$ and $\tilde{Z}\of{\ell,r}$ on a lattice. However, the replica method can also be applied to QFTs in a continuum formulation, provided suitable UV and IR regularization methods are available.}
    \label{fig:replica}
\end{figure}

For a partition of a Hilbert space into two regions $A$ and $B$, a reduced density matrix of $A$ can be constructed by tracing over the degrees of freedom of $B$, $\rho_A=\tra_B \rho$, where $\rho$ is the density matrix corresponding to a pure state of the Hilbert space of the whole system. The EE associated with $A$ is then defined as the von Neumann entropy of $\rho_A$
\be \label{eq:EE_def}
 \SEE(A) = -\tr\left(\rho_A\log \rho_A\right) \ .
\ee
While the definition is conceptually simple, due to the logarithm, EE is notoriously difficult to compute in interacting QFTs, requiring sophisticated techniques such as replica methods~\cite{Calabrese:2004eu,Calabrese:2009qy}.
The replica method relates a trace of powers of the reduced density matrix to a ratio of partition functions 
\begin{equation} \label{eq:replicatrick}
    \text{tr}(\rho^r_A)=\frac{\tilde{Z}(\ell,r)}{Z^r}\ .
\end{equation}
Here $Z$ is the standard finite temperature partition function with fields being $1/T$ periodic in the Euclidean time direction (cf. figure~\ref{fig:replica}, left). For $\tilde{Z}(\ell,r)$, the temporal periodicity is different in the two regions $A$ and $B$. In $A$, which we consider to be a slab of width $\ell$, the fields are $r/T$ periodic. In $B$, Euclidean time extent splits into $r$ disjoint domains, in each of which the fields are $1/T$ periodic (cf. figure~\ref{fig:replica}, right). Note that for $\rho$ to be a pure state, \eqref{eq:replicatrick} should a priori be considered in the limit $T\to 0$.  
Entanglement entropy can then be expressed in terms of the two partition functions
\begin{align} \label{EE}
        \SEE(\ell) & =-\lim_{r\rightarrow1}\frac{\partial\log\text{tr}(\rho^r_A)}{\partial r} \nonumber \\
         & =-\lim_{r\rightarrow1}\left(\frac{\partial\log \tilde{Z}(\ell,r)}{\partial r}-\log Z\right) \ .
\end{align}
By means of \eqref{EE}, entanglement measures are defined purely through path integrals, avoiding any explicit tensor-product decomposition of the Hilbert space, which is particularly advantageous in gauge field theories where such a decomposition is delicate due to gauge constraints~\cite{Casini:2013rba}.

At non-zero temperature $T$, the system is not in a pure state and the definition of EE given in \eqref{eq:EE_def} is in principle not applicable since the operator has overlap with thermal entropy. Consequently, \eqref{EE} will no longer measure purely the entropy of entanglement between regions $A$ and $B$, but also the thermal entropy of $A$, as discussed more extensively in the next section. Nevertheless, it is customary to refer to \eqref{EE} as \emph{entanglement entropy} even in the finite temperature context~\cite{Calabrese:2004eu,Calabrese:2009qy}, which we will also do.   

In strongly coupled QFTs, EE is a UV-divergent quantity, due to the strong short distance correlations across the boundary $\partial A$ between $A$ and $B$, with the divergence being proportional to the area of $\partial A$ \cite{Srednicki:1993im,Eisert:2008ur}. However, with a slab shaped entangling region as the one we introduced above to describe the replica trick, the derivative of EE with respect to $\ell$ is UV-finite.

\section{Entanglement thermodynamics} \label{sect:EEthermo}
In this section, we wish to connect entanglement and thermal entropy in the limit of large $\ell$. For $\ell \to\infty$, this can be seen directly from the definition \eqref{eq:EE_def}. In the limiting case, where $A$ encompasses the whole system and $B$ is empty, $\rho_A\to\rho$ and 
\begin{equation} \label{eq:EESth relation}
    \lim_{B\to\emptyset} \SEE\of{A}= -\text{tr}\left(\rho\log (\rho)\right)=S \ .
\end{equation}
At zero temperature, $S=0$ and \eqref{eq:EESth relation} is trivial, whereas at finite temperature, the relation reflects the aforementioned fact that the thermal system cannot be described by a pure state and \eqref{eq:EE_def} has overlap with the operator measuring the thermal entropy of $A$. As such, if $A$ is entangled with nothing, the entanglement entropy reduces to its thermal entropy $S$~\cite{Nishioka:2009un}.

A similar situation occurs, if we assume that both, $A$ and $B$ are finite\footnote{The regions are considered finite here to obtain finite values for the extensive entropies.} and that their linear extent in all directions is much larger than the largest correlation length $\xi$ of the theory. Then a large subpart $\tilde{A}$ of $A$ cannot be entangled with $B$ since $B$ is out of reach. For this subpart $\tilde{A}\subset A$, equation \eqref{eq:EESth relation} holds. Now, as thermal entropy is an extensive quantity, the thermal entropy of $\tilde{A}$ is proportional to the volume $\mathrm{vol}\ssof{\tilde{A}}$ of $\tilde{A}$. For a slab-shaped entangling region of width $\ell\gg\xi$, changing $\ell$ can be assumed to simply change $\mathrm{vol}\ssof{\tilde{A}}$ while $A\setminus\tilde{A}$, \ie, the part of $A$ that is within correlation range of region $B$, remains unaffected. Taking a partial derivative with respect to $\ell$ of the relation \eqref{eq:EESth relation} therefore yields
\begin{equation} \label{eq:dEEsth relation}
    \lim_{\xi\ll \ell}\frac{\partial \SEE(A)}{\partial \ell}=\frac{\partial\, \text{vol}(\tilde{A})}{\partial \ell} s \ ,
\end{equation}
where $s$ is the thermal entropy density in $\tilde{A}$ and $\partial_{\ell} \, \text{vol}(\tilde{A})=V_{\perp}$ is the area of the cross-section of $A$ perpendicular to the direction along which $\ell$ is aligned. Multiplying the volume factor to the left-hand side we have
\begin{equation} \label{eq:dEEsth relation lattice}
    \lim_{\xi\ll \ell}\frac{1}{V_{\perp}}\frac{\partial \SEE(A)}{\partial \ell}= s
\end{equation}
as the final form of the relation. Next, we will lay out a more rigorous derivation for this connection from the replica trick expression of entanglement entropy and introduce some thermodynamic relations which we can use to test \eqref{eq:dEEsth relation lattice} with lattice simulations. 

The following discussion refers to grand canonical ensembles, but can easily be carried over to the canonical case as well.

Let us consider the dimensionless free energy density $\omega\of{\beta,\mu}$ of a system described by some grand canonical partition function $Z\of{\beta,V,\mu}$, \ie, 
\begin{equation}
\omega\of{\beta,\mu}=-\lim_{V\to\infty}\frac{1}{V}\log\of{Z\of{\beta,V,\mu}}\ ,\label{eq:latfe}
\end{equation}
where $\beta$ is the inverse temperature and temporal extent, $\mu$ a chemical potential that couples to a conserved charge, and $V=\vspa$ the $(d-1)$-dimensional spatial volume with $L$ corresponding to the linear system size in the direction along which $\ell$ is aligned and $L_\perp$ the linear system size in the perpendicular directions.
The dimensionless free energy density \eqref{eq:latfe} is related to the Landau free energy density
\begin{equation}
\omega_L\of{T,\mu}=\epsilon-T\,s-\mu\,n=-p\ ,\label{eq:landaufe}
\end{equation} 
with $\epsilon$ being the internal energy density, $T=1/\beta$ the temperature, $s$ the thermal entropy density, and $n$ some charge density, by
\begin{equation}
\omega\of{\beta,\mu}=\beta\,\omega_L\of{T\of{\beta}, \mu}\ .\label{eq:latferel}
\end{equation}
The differential of $\omega_L\of{T,\mu}$ reads
\begin{equation}
\dd\omega_L=-s\,\dd T - n\,\dd\mu\ ,\label{eq:landaufediff}
\end{equation}
and, correspondingly, using $\omega_L=\omega/\beta$, the differential of $\omega\of{\beta,\mu}$ is given by
\begin{equation}
\dd\omega=\frac{\omega+s}{\beta}\,\dd \beta - \beta\,n\,\dd\mu\ .\label{eq:latfediff}
\end{equation}
From \eqref{eq:latfediff} it then follows that the thermal entropy can be obtained as
\begin{equation}
s=\beta\,\partdf{\omega}{\beta}{\mu}-\omega\ .\label{eq:thermsdensfromlatfe}
\end{equation}

\begin{figure}[ht]
    \centering
\includegraphics[width=0.97\columnwidth]{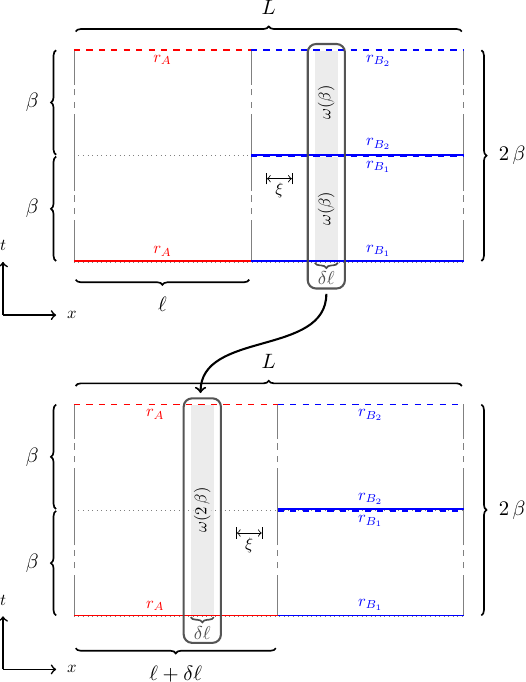}
\caption{Illustration of how the dimensionless free energy \eqref{eq:latlandauferep} changes in a replicated system with $r=2$ and an entangling region of width $\ell$. The variation corresponds to increasing $\ell$ by an infinitesimal amount $\delta\ell$ at fixed total size $L$ in the regime $\xi\ll\ell\ll L$, where $\xi$ denotes the longest correlation length. The red and blue horizontal lines illustrate the different temporal boundary conditions over the entangling region $A$ and its complement $B$: dashed and solid lines with the same label are identified. The process of increasing $\ell$ by $\delta\ell$ can be thought of as removing a slice of width $\delta\ell$ from deep within the bulk of region $B$ while adding a slice of width $\delta\ell$ to the bulk of region $A$. Since we assume that $\xi\ll\ell\ll L$, the bulk properties over region $A$ and $B$ are described by the free energy density \eqref{eq:latfe} for unreplicated systems, with inverse temperatures $r\,\beta$ and $\beta$, respectively.}
\label{fig:lvstempderivequiv}
\end{figure}

To make contact with entanglement, recall that entanglement entropy can be expressed as
\begin{multline}
\SEE\of{\ell}=\lim_{r\to 1}\partd{}{r}\sof{\tilde{\Omega}\of{\ell,\beta,V,\mu,r} - r\,\Omega\of{\beta,V,\mu}}\\
=\lim_{r\to 1}\bof{\partd{\tilde{\Omega}\of{\ell,\beta,V,\mu,r}}{r} - \Omega\of{\beta,V,\mu}}\ ,\label{eq:seedef}
\end{multline}
where
\begin{equation}
\tilde{\Omega}\of{\ell,\beta,V,\mu,r}=-\log\of{\tilde{Z}\of{\ell,\beta,V,\mu,r}}\ ,\label{eq:latlandauferep}
\end{equation}
is the dimensionless free energy of a system with $r$ replicas, each with temporal extent $\beta$, and total spatial volume $V=\vspa$, and in which the entangling region $A$ has width $\ell$, and 
\begin{equation}
\Omega\of{\beta,V,\mu}=-\log\of{Z\of{\beta,V,\mu}}\ \label{eq:latlaundaufehomog}
\end{equation}
is the dimensionless free energy of a homogeneous system (without replicas) of temporal extent $\beta$ and spatial volume $V$. With \eqref{eq:seedef} the $\ell$-derivative of EE scaled by $1/V_\perp=1/L_\perp^{d-2}$ reads
\begin{multline}
\frac{1}{V_\perp}\partdf{\SEE\of{\ell}}{\ell}{\beta,V,\mu}=\\
\frac{1}{V_\perp}\lim_{r\to 1}\partd{}{\ell}\bof{\partd{\tilde{\Omega}\of{\ell,\beta,V,\mu,r}}{r} - \Omega\of{\beta,V,\mu}}\\
=\frac{1}{V_\perp}\lim_{r\to 1} \spartd{\tilde{\Omega}\of{\ell,\beta,V,\mu,r}}{\ell}{r}\\
=\lim_{r\to 1}\partd{}{r} \bof{\frac{1}{V_\perp}\partd{\tilde{\Omega}\of{\ell,\beta,V,\mu,r}}{\ell}}\ ,\label{eq:seelsderiv}
\end{multline}
where we assumed after the first equality sign that taking the limit $\of{r\to 1}$ commutes with taking the $\ell$-derivative.

Next, following the argument from~\cite{Jokela:2023rba}, we write the $\ell$-derivative in \eqref{eq:seelsderiv} as 
\begin{multline}
    \partd{\tilde{\Omega}\of{\ell,\beta,V,\mu,r}}{\ell}=\\
    \frac{\tilde{\Omega}\of{\ell+\delta\ell,\beta,V,\mu,r}-\tilde{\Omega}\of{\ell,\beta,V,\mu,r}}{\delta\ell}\ . \label{eq:lderivative}
\end{multline}
The increase in $\ell$ by $\delta\ell$ in the numerator can be thought of as removing a slice of width $\delta\ell$ from the complement $B$ and adding it to the entangling region $A$ with the corresponding change in the temporal boundary conditions as shown in figure~\ref{fig:lvstempderivequiv}. Let us consider a case, where the slice is taken from deep within $B$ and placed deep within $A$, \ie, inside $\tilde{A}$. For $\xi\ll\ell\ll L$, \ie, if the linear sizes of $A$ and $B$ are both much larger than the longest correlation length $\xi$ of the theory described by \eqref{eq:latlaundaufehomog}, the effect of the interface between the two regions can be neglected when describing thermodynamics deep inside either region. Then, the difference of the dimensionless free energies in the numerator of \eqref{eq:lderivative} reduces to just the difference between the dimensionless free energies of the slice of width $\delta\ell$ that is added to $A$ and removed from $B$. Multiplying the expression by $1/V_\perp$, the denominator becomes the spatial volume of the slice, and we are left with a difference between dimensionless free energy densities. In the limit considered, deep inside $B$, the system consists of $r$ replica, each of which is described by the dimensionless free energy density $\omega\of{\beta,\mu}$, and deep inside $A$, the system consists of just one part described by the dimensionless free energy density $\omega\of{r\beta,\mu}$. This
leads to the following expression:
\begin{multline}
\lim_{\underset{\ell\ll L}{\ell,L\to\infty}}\frac{1}{V_\perp}\partdf{\tilde{\Omega}\of{\ell,\beta,V,\mu,r}}{\ell}{\beta,V,\mu,r}=\\
\omega\of{r\,\beta,\mu} - r\,\omega\of{\beta,\mu}\ .\label{eq:felatlargelapprox}
\end{multline}
By combining \eqref{eq:seelsderiv} with \eqref{eq:felatlargelapprox} we obtain
\begin{multline}
\lim_{\underset{\ell\ll L}{\ell,L\to\infty}}\frac{1}{V_\perp}\partdf{\SEE\of{\ell}}{\ell}{\beta,V,\mu}=\\
\lim_{r\to 1}\partd{}{r} \sof{\omega\of{r\,\beta,\mu} - r\,\omega\of{\beta,\mu}}\\
=\beta\,\partd{\omega\of{\beta,\mu}}{\beta} - \omega\of{\beta,\mu}\ ,
\end{multline}
and therefore find with \eqref{eq:thermsdensfromlatfe} that
\begin{equation}
\lim_{\underset{\ell\ll L}{\ell,L\to\infty}}\frac{1}{V_\perp}\partdf{\SEE\of{\ell}}{\ell}{\mathrlap{\beta,V,\mu}}\,\,\,\,=\beta\,\partdf{\omega}{\beta}{\mu}-\omega=s\ .\label{eq:eederiveqtherms}
\end{equation}

Thus, for $\xi\ll\ell\ll L$, the $\ell$-derivative of $\SEE\of{\ell}/V_\perp$ is equivalent to the negative of the $T$-derivative of the Landau free energy density, Eq.~\eqref{eq:landaufe}, since it follows from \eqref{eq:landaufediff} that
\begin{equation}
s=-\partdf{\omega_L}{T}{\mu}\ .\label{eq:thermsdensfromlandaufe}
\end{equation}
Note that in \eqref{eq:eederiveqtherms} we could substitute 
\begin{equation}
\frac{1}{V_\perp}\partdf{}{\ell}{\beta,V,\mu} \rightarrow \partdf{}{\text{vol}\of{A}}{\beta,V,\text{shape}\of{\partial A},\mu} \ ,
\end{equation}
where $\text{vol}\of{A}$ is the spatial volume of the entangling region $A$, and $\partial A$ its boundary, the entangling surface. The derivative with respect to the entangling region volume, $\text{vol}\of{A}$, needs to be taken with not only the total volume, $V$, but also the shape and size of the entangling surface $\partial A$ kept fixed.

\iftrue
By taking the derivative with respect to $\mu$ on both sides of \eqref{eq:eederiveqtherms} and using the Maxwell relation 
\begin{equation}
\partdf{s}{\mu}{T}=\partdf{n}{T}{\mu}\ ,
\end{equation}
we find that
\begin{equation}
\frac{1}{V_\perp}\frac{\partial^2\SEE}{\partial\mu\,\partial\ell}\Bigg|_{\beta,V,\mu}=-\beta^2\,\partdf{n}{\beta}{\mu}\ \ , \ \  \xi\ll\ell\ll L \ .\label{eq:thermsdensmaxrel}
\end{equation}

In lattice computations, it is customary to approximate $\SEE$ by $H_2$, where $H_r$ refers to the $r$-th R\'{e}nyi entropy,
\begin{multline}
H_r\of{\ell,\beta,V,\mu}=\frac{1}{1-r}\log\trace\of{\rho_A^r}\\
=\frac{-1}{1-r}\of{\tilde{\Omega}\of{\ell,\beta,V,\mu,r}-r\,\Omega\of{\beta,V,\mu}}\ ,\label{eq:renyientropy}
\end{multline}
for which one has:
\begin{equation}\label{eq:HrlogZrelation}
\frac{\partial H_r\of{\ell}}{\partial \ell} = \frac{-1}{1-r}\frac{\partial\tilde{\Omega}\of{\ell,\beta,V,\mu,r}}{\partial \ell}\ .
\end{equation}

We can also derive relations analogous to \eqref{eq:eederiveqtherms} and \eqref{eq:thermsdensmaxrel} for the R\'enyi entropies of integer order $r\geq 2$. It follows from \eqref{eq:felatlargelapprox} that
\begin{equation}
\lim_{\underset{\ell\ll L}{\ell,L\to\infty}}\frac{1}{V_{\perp}}\frac{\partial H_r\of{\ell}}{\partial \ell}=s_{r}\of{T\of{\beta},\mu}\ ,\label{eq:hromegaexpr}
\end{equation}
with
\begin{multline}
s_r\of{T,\mu}=-\frac{\omega_L\of{T,\mu}-\omega_L\of{T/r,\mu}}{T-T/r}\\
\equiv -\Delta_{T}^{r}\,\omega_L\of{T,\mu}\ ,\label{eq:stepscaleentropydens}
\end{multline}
where we used $\omega\of{\beta,\mu}=\beta\,\omega_L\of{1/\beta,\mu}$ from \eqref{eq:latferel} and $T=1/\beta$. We also defined the \emph{step scaling derivative} with respect to $T$ for a scaling factor $r$. It provides a discrete approximation of the partial derivative with respect to $T$ in terms of a \emph{step scaling function} with scaling factor $r$. When extending $r$ to $\mathbb{R}_{\geq 1}$ one has 
\begin{equation}
\lim_{r\to 1} s_r\of{T,\mu}=s\of{T,\mu}\ ,\label{eq:stepscaleentropydenslimit}
\end{equation}
which is compatible with the fact that 
\begin{equation}
\SEE\of{\ell}=\lim_{r\to 1} H_r\of{\ell}\, , 
\end{equation}
so that \eqref{eq:hromegaexpr} becomes \eqref{eq:eederiveqtherms} in said limit.

By taking a partial derivative with respect to $\mu$ on both sides of \eqref{eq:hromegaexpr} and recalling that $n=-\partial_{\mu}\omega_L\of{T,\mu}$, one obtains
\begin{equation}
\frac{1}{V_{\perp}}\frac{\partial^2 H_r}{\partial\mu\,\partial\ell}=\Delta_{T}^{r}\,n\of{T,\mu}|_{T=T\of{\beta}}\ ,\ \ \xi\ll\ell\ll L\ ,\label{eq:maxwell_rel_hr}
\end{equation}
where the right-hand side,
\begin{multline}
\Delta_{T}^{r}\,n\of{T,\mu}=\frac{n\of{T,\mu}-n\of{T/r,\mu}}{T-T/r}\\
=\frac{n\of{\beta,\mu}-n\of{r\,\beta,\mu}}{1/\beta-1/\of{r\,\beta}}=-r\,\beta\,\frac{n\of{r\,\beta,\mu}-n\of{\beta,\mu}}{r-1}\\
=-r\,\beta^2\,\frac{n\of{r\,\beta,\mu}-n\of{\beta,\mu}}{r\,\beta-\beta}\label{eq:stepscalederivofdens}
\end{multline}
is again a discrete step scaling approximation of the right-hand side of \eqref{eq:thermsdensmaxrel} with scaling factor $r$, \ie, 
\begin{equation}\label{eq:nstepscalederiv}
\lim_{r\to 1}\Delta_{T}^{r}\,n\of{T,\mu} = \partdf{n}{T}{\mu}=-\beta^2\,\partdf{n}{\beta}{\mu}\ ,
\end{equation}
so that in the limit $r\to 1$ equation \eqref{eq:maxwell_rel_hr} reduces to \eqref{eq:thermsdensmaxrel}.

\fi

\section{$\On{N}$ models at finite density on the lattice}\label{sec:latticesetup}

Until this point, our considerations of entanglement and R\'enyi entropies have been for general QFTs. We now specialize to $\On{N}$ models, the specific theories of our lattice simulations. Although simple, $\On{N}$ models can have several interesting properties such as asymptotic freedom and spontaneous symmetry breaking, making them a fertile testing ground for new ideas. Additionally, on the lattice, they can be simulated directly at finite densities with the use of a worm algorithm~\cite{Prokofev:2001ddj}, unlike, \eg, QCD where complicated extrapolation techniques like Taylor expansions or analytic continuations have to be used \cite{Philipsen:2007rj,deForcrand:2009zkb}. This, along with their relatively low simulation cost, makes $\On{N}$ models ideal for studying EE at finite density.

In this section, we review the lattice formulation and sampling algorithm for $\On{N}$ models at finite density. This is done in preparation for the discussion in section~\ref{sect:latt_EE}, where the algorithms for measuring $\dHn{2}$ in this formalism are described.

\subsection{Continuum and lattice actions}
We consider the $\On{N}$ model with a $\phi_0^4$ interaction, a finite chemical potential $\mu_0$, and a source $J_0$. The subscript `0' is meant to indicate that these are continuum theory quantities. In general $d$ Euclidean spacetime dimensions, the continuum action of the theory is 
\begin{multline}
S\fof{\phi}\,=\,\int\text{d}^dx_0\bcof{\frac{1}{2}\left[\left(\partial^\nu+\,\mu_0\,\delta^{\nu}_{d}\,\tau_{12}\right)\phi_0\right]^{\intercal} \\
\cdot\left(\partial_\nu+\,\mu_0\,\delta_{\nu,d}\,\tau_{12}\right)\phi_0\\
+\frac{1}{2}m_0^2\,|\phi_0|^2+\frac{g_0}{4!}|\phi_0|^4+J_0\cdot\phi_0}\ ,\label{eq:onactioncont}
\end{multline}
where $\tau_{12}$ is an $N \times N$ matrix
\begin{equation}
    \tau_{12}=\begin{pmatrix} 
    0 & i & 0 & \cdots & 0 \\ 
    -i & 0 & 0 & \cdots & 0 \\ 
    0 & 0 &   & &  \vdots\\ 
    \vdots & & & \ddots &\\ 
    0 & \cdots & & & 0
    \end{pmatrix} \ .
\end{equation}
To obtain the lattice formulation of \eqref{eq:onactioncont}, we replace the integral with a sum and the partial derivatives with finite differences, following the standard procedure. Dimensionful quantities are also multiplied by suitable powers of the lattice spacing $a$ to obtain corresponding dimensionless quantities:
\begin{align}
    x_\nu&=\frac{x^\nu_0}{a} \label{eq:xlattunits} \\
    m&=a\,m_0 \label{eq:mlattunits}\\
    \mu&=a\mu_0 \label{eq:mulattunits} \\
    \sqrt{\kappa}\,\phi&=a^{(d-2)/2}\,\phi_0 \label{eq:philattunits} \\
    J&=a^{(d+2)/2}\,J_0 \label{eq:Jlattunits} \\
    g&=a^{4-d}\,g_0\ , \label{eq:glattunits}
\end{align}
where we introduced the hopping parameter $\kappa$. With these definitions, the action becomes:
\begin{multline}
S\,=\,\sum\limits_{x}\bcof{\frac{\kappa}{2}\,\sum\limits_{\nu=1}^{d}\sof{\of{\phi^{\intercal}_{x+\hat{\nu}}-\phi^{\intercal}_{x}}-\delta_{\nu,d}\,\mu\,\phi_{x}^{\intercal}\,\tau_{12}} \\
\cdot\sof{\of{\phi_{x+\hat{\nu}}-\phi_{x}}+\delta_{\nu,d}\,\mu\,\tau_{12}\,\phi_{x}}\\
+ \frac{\kappa\,m^2}{2} \abs{\phi_{x}}^{2} + \frac{\kappa^2\,g}{4!}\abs{\phi_{x}}^{4} - \sqrt{\kappa}\,\of{J\cdot\phi_{x}}}\\
=\,\sum\limits_{x}\bcof{-\frac{\kappa}{2}\,\sum\limits_{\nu=1}^{d}\sof{\phi^{\intercal}_{x}\,\of{1+\delta_{\nu,d}\,\mu\,\tau_{12}}\,\phi_{x+\hat{\nu}}\\
+\phi^{\intercal}_{x+\hat{\nu}}\,\of{1-\delta_{\nu,d}\,\mu\,\tau_{12}}\,\phi_{x}+\phi^{\intercal}_x\sof{\delta_{\nu,d}\,\mu^2-2}\phi_x}\\
+ \frac{\kappa\,m^2}{2} \abs{\phi_{x}}^{2} + \frac{\kappa^2\,g}{4!}\abs{\phi_{x}}^{4} - \sqrt{\kappa}\,\of{J\cdot\phi_{x}}}\\
=\,\sum\limits_{x}\bcof{-\frac{\kappa}{2}\,\sum\limits_{\nu=1}^{d}\sof{\phi^{\intercal}_{x}\,\e^{\mu\,\tau_{12}\,\delta_{\nu,d}}\,\phi_{x+\hat{\nu}}\\
+\phi^{\intercal}_{x+\hat{\nu}}\,\e^{-\mu\,\tau_{12}\,\delta_{\nu,d}}\,\phi_{x}+\,\sorder{a^3}}\\
+ \frac{\kappa}{2}\,\sof{2\,d+m^2} \abs{\phi_{x}}^{2} + \frac{\kappa^2\,g}{4!}\abs{\phi_{x}}^{4} - \sqrt{\kappa}\,\of{ J\cdot\phi_{x}}}\ .\label{eq:phifouractioncontinuumtolat}
\end{multline}
Before the last equality sign, the fact that $\phi_{x+\hat{\nu}}=\phi_{x}+(a\,\partial_{\nu}\,\phi_{x}) + \mathcal{O}(a^{2})$ was used, so that
\begin{multline}
\delta_{\nu,d}\,\mu^2\,\abs{\phi_x}^2 = \delta_{\nu,d}\,\mu^2\,\phi_x^{\intercal}\phi_x =\\
\frac{1}{2}\delta_{\nu,d}\,\mu^2\,(\phi^{\intercal}_x\,\phi_{x+\hat{\nu}} + \phi^{\intercal}_{x+\hat{\nu}}\,\phi_x + \underbrace{\mathcal{O}(a)}_{\mathclap{-\phi^{\intercal}_x\,(a\,\partial_{\nu}\phi_{x})-(a\,\partial_{\nu}\phi^{\intercal}_{x})\,\phi_x}})\\
= \frac{1}{2}\delta_{\nu,d}\,\mu^2\,(\phi^{\intercal}_x\,\phi_{x+\hat{\nu}} + \phi^{\intercal}_{x+\hat{\nu}}\,\phi_x) + \mathcal{O}(a^{3}),
\end{multline}
and therefore
\begin{multline}
\phi^{\intercal}_x\,(1+\delta_{\nu,d}\,\mu\,\tau_{12})\,\phi_{x+\hat{\nu}} + \phi^{\intercal}_{x+\hat{\nu}}\,(1-\delta_{\nu,d}\,\mu\,\tau_{12})\,\phi_x\\
+ \delta_{\nu,d}\,\mu^2\,\abs{\phi_x}^2\\
=\phi^{\intercal}_x\,(1+\delta_{\nu,d}\,\mu\,\tau_{12}+\frac{(\delta_{\nu,d}\,\mu)^2}{2})\,\phi_{x+\hat{\nu}}\\
+ \phi^{\intercal}_{x+\hat{\nu}}\,(1-\delta_{\nu,d}\,\mu\,\tau_{12}+\frac{(\delta_{\nu,d}\,\mu)^2}{2})\,\phi_x + \mathcal{O}(a^{3})\\
=\phi^{\intercal}_x\,e^{\mu\,\tau_{12}\,\delta_{\nu,d}}\,\phi_{x+\hat{\nu}} + \phi^{\intercal}_{x+\hat{\nu}}\,e^{-\mu\,\tau_{12}\,\delta_{\nu,d}}\,\phi_x + \mathcal{O}(a^{3})\ .
\end{multline}
Finally, with the following definitions,
\begin{align}
    \lambda&=\frac{\kappa^2\,g}{4!} \\
    m^2&=\frac{2-4\lambda}{\kappa}-2d \\
    j&=\sqrt{\kappa} J \ ,
\end{align}
the $\On{N}$ lattice action with a chemical potential can be written in the form
\begin{multline}
S\fof{\phi}\,=\,\sum\limits_{x}\bcof{-\frac{\kappa}{2}\,\sum\limits_{\nu=1}^{d}\sof{\phi^{\intercal}_{x}\,\e^{\mu\,\tau_{12}\,\delta_{\nu,d}}\,\phi_{x+\hat{\nu}}\\
+\phi^{\intercal}_{x+\hat{\nu}}\,\e^{-\mu\,\tau_{12}\,\delta_{\nu,d}}\,\phi_{x}}\\
\qquad +\abs{\phi_x}^2+\lambda\ssof{\abs{\phi_x}^2-1}^{2}-j\cdot\phi_{x}}\ ,\label{eq:onaction}
\end{multline}
which allows for easy switching between linear and non-linear $\On{N}$ models, since the partition function of the latter can be obtained from the partition function $Z=\int\mathcal{D}[\phi]\e^{-S[\phi]}$ with \eqref{eq:onaction} by formally taking the limit $\lambda\to\infty$.

\subsection{Dual variable representation of the partition function}

The action \eqref{eq:onaction} of the $\On{N}$ model becomes complex if $\mu\neq0$. Therefore, at $\mu\neq0$, $\mathcal{D}[\phi]\e^{-S[\phi]}$ has no probabilistic meaning and importance sampling cannot be done in terms of the fields $\phi$. To circumvent this sign problem, we reformulate the theory in terms of integer-valued dual variables which can be sampled with a worm algorithm \cite{Prokofev:2001ddj}. There are multiple ways to formulate the system in terms of dual variables~\cite{Endres:2006xu,Gattringer:2012df,Gattringer:2012ap,Bruckmann:2015sua,Katz:2016azl}. We use the formalism from \cite{Rindlisbacher:2015xku,Rindlisbacher:2016zht,Rindlisbacher:2017ysn} which we will describe in the following subsections.

We start by considering the partition function $Z=\int\mathcal{D}[\phi]\e^{-S[\phi]}$ and writing the exponential of the action \eqref{eq:onaction} as a product of individual exponentials
\begin{multline}
Z\,=\,\int\DD{\phi}\prod\limits_{x}\bcof{\bof{\prod\limits_{\nu=1}^{d}\exp\sof{\kappa\,\e^{\mu\,\delta_{\nu,d}}\,\phi^{+}_{x}\,\phi^{-}_{x+\hat{\nu}}}\\
\cdot\exp\sof{\kappa\,\e^{-\mu\,\delta_{\nu,d}}\,\phi^{+}_{x+\hat{\nu}}\,\phi^{-}_{x}}\,\bof{\prod\limits_{i=3}^{N}\,\exp\sof{\kappa\,\phi^{i}_{x}\,\phi^{i}_{x+\hat{\nu}}}}}\\
\cdot\exp\sof{-\abs{\phi_{x}}^2-\lambda\sof{\abs{\phi_{x}}^{2}-1}^{2}}\\
\cdot\exp\sof{j^{+}\,\phi^{+}_{x}}\exp\sof{j^{-}\,\phi^{-}_{x}}\,\bof{\prod\limits_{i=3}^{N}\exp\sof{j^{i}\,\phi^{i}_{x}}}}\ .\label{eq:phifourfluxrepd1}
\end{multline}
Here we introduced $\phi^{\pm}=\frac{1}{\sqrt{2}}\of{\phi_{1}\pm\ii\,\phi_{2}}$ and $j^{\pm}=\frac{1}{\sqrt{2}}\of{j_{1}\mp\ii\,j_{2}}$ to diagonalize $\tau_{12}$. All of the exponentials, except the ones corresponding to the potential term, are then written as power series,
\begin{multline}
\exp\sof{\kappa\,\e^{\mu\delta_{\nu,d}}\,\phi^{+}_{x}\,\phi^{-}_{x+\hat{\nu}}}\,=\\
\sum\limits_{\eta_{x,\nu}}\frac{\sof{\phi^{+}_{x}\,\phi^{-}_{x+\hat{\nu}}\,\kappa\,\e^{\mu\delta_{\nu,d}}}^{\eta_{x,\nu}}}{\eta_{x,\nu}!}
\end{multline}
\begin{multline}
\exp\sof{\kappa\,\e^{-\mu\delta_{\nu,d}}\,\phi^{+}_{x+\hat{\nu}}\,\phi^{-}_{x}}\,=\\
\sum\limits_{\bar{\eta}_{x,\nu}}\frac{\sof{\phi^{+}_{x+\hat{\nu}}\,\phi^{-}_{x}\,\kappa\,\e^{-\mu\delta_{\nu,d}}}^{\bar{\eta}_{x,\nu}}}{\bar{\eta}_{x,\nu}!}
\end{multline}
\begin{equation}
\exp\sof{\kappa\,\phi^{i}_{x}\,\phi^{i}_{x+\hat{\nu}}}\,=\,\sum\limits_{\chi^{\of{i}}_{x,\nu}}\frac{\sof{\kappa\,\phi^{i}_{x}\,\phi^{i}_{x+\hat{\nu}}}^{\chi^{\of{i}}_{x,\nu}}}{\chi^{\of{i}}_{x,\nu}!}\ ,
\end{equation}
and
\begin{equation} \label{eq:s+pow.ser.}
\exp\of{j^{+}\,\phi^{+}_{x}}\,=\,\sum\limits_{m_{x}}\frac{\sof{j^{+}\,\phi^{+}_{x}}^{m_{x}}}{m_{x}!}
\end{equation}
\begin{equation} \label{eq:smpow.ser.}
\exp\of{j^{-}\,\phi^{-}_{x}}\,=\,\sum\limits_{\bar{m}_{x}}\frac{\sof{j^{-}\,\phi^{-}_{x}}^{\bar{m}_{x}}}{\bar{m}_{x}!}
\end{equation}
\begin{equation} \label{eq:sipow.ser.}
\exp\of{j^{i}\,\phi^{i}_{x}}\,=\,\sum\limits_{n^{\of{i}}_{x}}\frac{\sof{j^{i}\,\phi^{i}_{x}}^{n^{\of{i}}_{x}}}{n^{\of{i}}_{x}!}\ .
\end{equation}
Next, the $\phi_{x}$ are expressed in polar form, \ie,
\begin{subequations}
\begin{align}
\phi^{+}_{x}\,&=\,S_{x}\,\sin\sof{\theta^{\of{N-1}}_{x}}\,\cdots\,\sin\sof{\theta^{\of{2}}_{x}}\,\frac{\e^{\ii\,\theta^{\of{1}}_{x}}}{\sqrt{2}}\\
\phi^{-}_{x}\,&=\,S_{x}\,\sin\sof{\theta^{\of{N-1}}_{x}}\,\cdots\,\sin\sof{\theta^{\of{2}}_{x}}\,\frac{\e^{-\ii\,\theta^{\of{1}}_{x}}}{\sqrt{2}}\\
\phi^{3}_{x}\,&=\,S_{x}\,\sin\sof{\theta^{\of{N-1}}_{x}}\,\cdots\,\cos\sof{\theta^{\of{2}}_{x}}\\
\vdots & \nonumber \\
\phi^{N-1}_{x}\,&=\,S_{x}\,\sin\sof{\theta^{\of{N-1}}_{x}}\,\cos\sof{\theta^{\of{N-2}}_{x}}\\
\phi^{N}_{x}\,&=\,S_{x}\,\cos\sof{\theta^{\of{N-1}}_{x}}\ ,
\end{align}
\end{subequations}
and accordingly the integration measure becomes
\begin{multline}
\DD{\phi}\,\propto\,\prod\limits_{x}\,\bof{S^{N-1}_{x}\,\bof{\prod\limits_{i=2}^{N-1}\,\sof{\sin\sof{\theta_{x}^{\of{i}}}}^{i-1}}\\
\dd{\theta_{x}^{\of{1}}}\wedge\cdots\wedge \dd{\theta_{x}^{\of{N-1}}}\wedge\dd{S_{x}}}\ .
\end{multline}
With these operations, the partition function \eqref{eq:phifourfluxrepd1} can be expressed as
\begin{widetext}
\begin{multline}
Z\,=\,\sum\limits_{\cof{\eta,\bar{\eta},m,\bar{m}}}\prod\limits_{x}\bcof{\bof{\prod\limits_{\nu}\frac{\of{\frac{\kappa}{2}}^{\eta_{x,\nu}+\bar{\eta}_{x,\nu}}}{\eta_{x,\nu}!\,\bar{\eta}_{x,\nu}!}}\frac{\sof{\frac{j^{+}}{\sqrt{2}}}^{m_{x}}\sof{\frac{j^{-}}{\sqrt{2}}}^{\bar{m}_{x}}}{m_{x}!\,\bar{m}_{x}!}\,\bof{\prod\limits_{i=3}^{N}\bof{\prod\limits_{\nu}\frac{\kappa^{\chi^{\of{i}}_{x,\nu}}}{\chi^{\of{i}}_{x,\nu}!}}\frac{\sof{j^{i}}^{n^{\of{i}}_{x}}}{n^{\of{i}}_{x}!}}\\
\cdot\e^{\mu\of{\eta_{x,d}-\bar{\eta}_{x,d}}}\,\int\limits_{-\pi}^{\pi}\idd{\theta_{x}^{\of{1}}}{}\,\e^{\ii\,\theta^{\of{1}}_{x}\,\sof{m_{x}-\bar{m}_{x}+\smash{\sum\limits_{\nu}}\of{\eta_{x,\nu}-\bar{\eta}_{x,\nu}-\of{\eta_{x-\hat{\nu},\nu}-\bar{\eta}_{x-\hat{\nu},\nu}}}}}\\
\cdot\bof{\prod\limits_{a=2}^{N-1}\,\int\limits_{0}^{\pi}\idd{\theta_{x}^{\of{a}}}{}\,\sof{\cos\ssof{\theta_{x}^{\of{a}}}}^{n^{\of{a+1}}_{x}+\sum\limits_{\nu}\sof{\chi^{\of{a+1}}_{x,\nu\vphantom{\hat{\nu}}}+\chi^{\of{a+1}}_{x-\hat{\nu},\nu}}}\\
\cdot\sof{\sin\ssof{\theta_{x}^{\of{a}}}}^{a-1+m_x+\bar{m}_x+\sum\limits_{i=3}^{a} n^{\of{i}}_{x}+\sum\limits_{\nu}\sof{\eta_{x,\nu}+\bar{\eta}_{x,\nu}+\eta_{x-\hat{\nu},\nu}+\bar{\eta}_{x-\hat{\nu},\nu}+\sum\limits_{i=3}^{a}\sof{\chi^{\of{i}}_{x,\nu\vphantom{\hat{\nu}}}+\chi^{\of{i}}_{x-\hat{\nu},\nu}}}}}\\
\cdot\int\limits_{0}^{\infty}\idd{S_{x}}{}\,S_{x}^{N-1+m_{x}+\bar{m}_{x}+\smash{\sum\limits_{i=3}^{N}} n^{\of{i}}_{x}+\smash{\sum\limits_{\nu}}\sof{\eta_{x,\nu}+\bar{\eta}_{x,\nu}+\eta_{x-\hat{\nu},\nu}+\bar{\eta}_{x-\hat{\nu},\nu}\,+\,\smash{\sum\limits_{i=3}^{N}}\sof{\chi^{\of{i}}_{x,\nu\vphantom{\hat{\nu}}}+\chi^{\of{i}}_{x-\hat{\nu},\nu}}}}\,\e^{-S_{x}^{2}-\lambda\of{S_{x}^{2}-1}^{2}}}\ .\label{eq:onfluxrepd2}
\end{multline}
\end{widetext}
Next, new \emph{flux variables} $k_{x,\nu}\in \mathbb{Z}$ and $l_{x,\nu}\in \mathbb{N}_{0}$ are defined such that
\begin{equation}
\eta_{x,\nu}-\bar{\eta}_{x,\nu}\,=\,k_{x,\nu}\quad,\quad \eta_{x,\nu}+\bar{\eta}_{x,\nu}\,=\,\abs{k_{x,\nu}}+2\,l_{x,\nu}\ .\label{eq:klvariables2}
\end{equation} 
Additionally, new \emph{monomer variables} $p_{x}\in \mathbb{Z}$ and $q_{x}\in \mathbb{N}_{0}$ are introduced such that
\begin{equation}
m_{x}-\bar{m}_{x}\,=\,p_{x}\quad,\quad m_{x}+\bar{m}_{x}\,=\,\abs{p_{x}}+2\,q_{x}\ .\label{eq:pqvariables2}
\end{equation}
We also define quantities
\begin{align}
A_{x}\,=&\,\sum\limits_{\nu}\of{\sabs{k_{x,\nu}}+\sabs{k_{x-\hat{\nu},\nu}}+2\sof{l_{x,\nu}+l_{x-\hat{\nu},\nu}}} \\
B^{i}_{x}\,=&\,\sum\limits_{\nu}\of{\chi^{\of{i}}_{x,\nu}+\chi^{\of{i}}_{x-\hat{\nu},\nu}}
\end{align}
at each site $x$ to shorten the notation. The angular integrals can be calculated analytically, using that for $M,N\,\in\,\mathbb{N}_{0}$:
\begin{multline}
\int\limits_{0}^{\pi}\idd{\theta}{}\,\sin^{M}\!\of{\theta}\,\cos^{N}\!\of{\theta}\,=\\
\frac{1+\of{-1}^{N}}{2}\frac{\Gamma\sof{\frac{1+M}{2}}\,\Gamma\sof{\frac{1+N}{2}}}{\Gamma\sof{\frac{2+M+N}{2}}}\ .
\end{multline}
The integrals over the $S_{x}$ variables cannot be performed analytically, but we can define corresponding weights
\begin{multline}
W_{\lambda,N}\of{n}\,=\,\int\limits_{0}^{\infty}\idd{S}{}\,S^{N-1+n}\,\e^{-S^{2}-\lambda\of{S^{2}-1}^{2}}\\
=\,\int\limits_{0}^{\infty}\idd{u}{}\,\frac{u^{\frac{n+N-2}{2}}\,\e^{-u-\lambda\of{u-1}^{2}}}{2}\ ,\label{eq:weightflambda}
\end{multline}
whose values can be computed numerically for any necessary $N$ and $\lambda$ at the beginning of a simulation.

Finally, with these definitions, the partition function \eqref{eq:onfluxrepd2} takes the form
\begin{multline}
Z = \sum\limits_{\mathclap{\cof{k,l,\chi,p,q,n}}}\quad\,\prod\limits_{x}\bcof{\bof{\prod\limits_{\nu=1}^{d}\frac{\of{\frac{\kappa}{2}}^{\abs{k_{x,\nu}}+2\,l_{x,\nu}}}{\of{\abs{k_{x,\nu}}+l_{x,\nu}}!\,l_{x,\nu}!}\bof{\prod\limits_{i=3}^{N}\frac{\kappa^{\chi^{\of{i}}_{x,\nu}}}{\chi^{\of{i}}_{x,\nu}!}}}\\
\cdot\frac{\sof{\frac{j^{+}}{\sqrt{2}}}^{\frac{1}{2}\of{\abs{p_{x}}+p_{x}}+q_{x}}\sof{\frac{j^{-}}{\sqrt{2}}}^{\frac{1}{2}\of{\abs{p_{x}}-p_{x}}+q_{x}}\,\e^{\mu\,k_{x,d}}}{\of{\abs{p_{x}}+q_{x}}!\,q_{x}!}\bof{\prod\limits_{i=3}^{N}\frac{\ssof{j^{i}}^{n^{\of{i}}_{x}}}{n^{\of{i}}_{x}!}}\\
\cdot\frac{\Gamma\sbof{\frac{2+A_{x}+\abs{p_{x}}+2\,q_{x}}{2}}\,\prod\limits_{i=3}^{N}\frac{1+\of{-1}^{B^{i}_{x}+n^{\of{i}}_{x}}}{2}\,\Gamma\sbof{\frac{1+B^{i}_{x}+n^{\of{i}}_{x}}{2}}}{\Gamma\sbof{\frac{N+A_{x}+\abs{p_{x}}+2\,q_{x}+\sum_{i=3}^{N}\ssof{B^{i}_{x}+n^{\of{i}}_{x}}}{2}}}\\
\cdot\delta\sof{p_{x}+\sum\limits_{\nu}\sof{k_{x,\nu}-k_{x-\hat{\nu},\nu}}}\\
\cdot W_{\lambda,N}\sof{A_{x}+\abs{p_{x}}+2\,q_{x}+\sum_{i=3}^{N}\ssof{B^{i}_{x}+n^{\of{i}}_{x}}}}\\
=\sum\limits_{\cof{k,l,\chi,p,q,n}}\prod\limits_{x}\bcof{\bof{\prod\limits_{\nu=1}^{d}\frac{\kappa^{\abs{k_{x,\nu}}+2\,l_{x,\nu}+\sum_{i=3}^{N}\,\chi^{\of{i}}_{x,\nu}}}{\of{\abs{k_{x,\nu}}+l_{x,\nu}}!\,l_{x,\nu}!\,\prod_{i=3}^{N}\chi^{\of{i}}_{x,\nu}!}}\\
\cdot\frac{j^{\abs{p_{x}}+2\,q_{x}}\,\e^{\ii\,\phi_{j}\,p_{x}}\,\e^{\mu\,k_{x,d}}}{2^{\sfrac{\ssof{\abs{p_{x}}+2\,q_{x}}}{2}}\of{\abs{p_{x}}+q_{x}}!\,q_{x}!}\bof{\prod\limits_{i=3}^{N}\frac{\ssof{j^{i}}^{n^{\of{i}}_{x}}}{n^{\of{i}}_{x}!}}\\
\cdot\delta\sof{p_{x}+\sum\limits_{\nu}\sof{k_{x,\nu}-k_{x-\hat{\nu},\nu}}}\\
\cdot W\sof{A_{x}+\abs{p_{x}}+2\,q_{x},\,B^{3}_{x}+n^{\of{3}}_{x},\,\ldots,\,B^{N}_{x}+n^{\of{N}}_{x}}}\ ,\label{eq:onfluxreppartf}
\end{multline}
where we introduced the discrete delta function $\delta\of{\cdot}=\delta_{\cdot,0}$ , and defined $j=\sqrt{j_{1}^{2}+j_{2}^{2}}$, $\phi_j=\arg\of{j_1-\ii\,j_2}$, and
\begin{multline}
W\sof{A,B^{3},\ldots,B^{N}}\,=\,\frac{\Gamma\sof{\frac{2+A}{2}} \prod\limits_{i=3}^{N}\frac{1+\of{-1}^{B^{i}}}{2}\Gamma\sof{\frac{1+\vphantom{B}\smash{B^{i}}}{2}}}{2^{\sfrac{A}{2}}\,\Gamma\sof{\frac{N+A+\sum_{i=3}^{N}B^{i}}{2}}}\\
\cdot W_{\lambda,N}\sof{A+\sum_{i=3}^{N}B^{i}}\ .\label{eq:onweightfunc0}
\end{multline}

After having defined the partition function in terms of the dual variables $k_{x,\nu},\,l_{x,\nu},\, \chi^{\of{i}}_{x,\nu},\,p_x,\, q_x$, and $n^{\of{i}}_x$, let us describe the physical meaning of these variables: $k_{x,\nu}$ is the net flux of charge from site $x$ to site $x+\hat{\nu}$. $l_{x,\nu}$ and $\chi^{\of{i}}_{x,\nu}$ are respectively the number of neutral $\phi^+\phi^-$ pairs and $\phi^i$ particles moving between the two sites. $p_x$, $q_x$, and $n^{\of{i}}_x$ count the total charge, the number of $\phi^+\phi^-$ pairs and $\phi^i$ particles at site $x$ respectively.

The discrete delta functions in \eqref{eq:onfluxreppartf} imply that $p_x=-\sum_{\nu}\sof{k_{x,\nu}-k_{x-\hat{\nu},\nu}}$. Combining this with the fact that $\sum_{x,\nu}\sof{k_{x,\nu}-k_{x-\hat{\nu},\nu}}=0$, since each $k$ variable appears in the sum twice with opposite signs, gives $\sum_x p_x=0$. This, in turn, indicates that $\prod_x \e^{\ii\,\phi_{j}\,p_{x}}=1$ meaning that our new partition function is free of any sign problems.

\subsection{Worm algorithm} \label{subsect:worm}

Despite the absence of a sign problem, there are some issues with the dual partition function \eqref{eq:onfluxreppartf}. At every site $x$, the expression in the delta function 
\begin{equation}\label{eq:deltafunctionconstraint}
\delta\sof{p_{x}+\sum_{\nu}\sof{k_{x,\nu}-k_{x-\hat{\nu},\nu}}}
\end{equation}
has to always be zero and the power in the \emph{evenness constraint},
\begin{equation}\label{eq:evennessconstraint}
1+\left(-1\right)^{\sum\limits_{\nu}\of{\chi^{\of{i}}_{x,\nu}+\chi^{\of{i}}_{x-\hat{\nu},\nu}}+n^{\of{i}}_x}
\end{equation}
has to always be even for the configuration to have a non-zero weight. If any of these two constraints are violated on a site $x$, we say that $x$ contains a \emph{defect}. 
Such major constraints on the allowed configurations of $k, \chi^{\of{i}}$, $p$, and $n^{\of{i}}$ variables mean that they cannot be sampled with regular local Metropolis updates.\footnote{The exceptions are the $l$ and $q$ variables, which are not subject to any constraints. The sum over $q$ in the partition function \eqref{eq:onfluxreppartf} can, however, be easily carried out numerically and absorbed in the pre-computed weights, removing the need to sample these variables~\cite{Rindlisbacher:2017ysn}.} They are instead sampled with the worm algorithm.

\begin{figure*}[ht]
    \centering
    \includegraphics[scale=0.68]{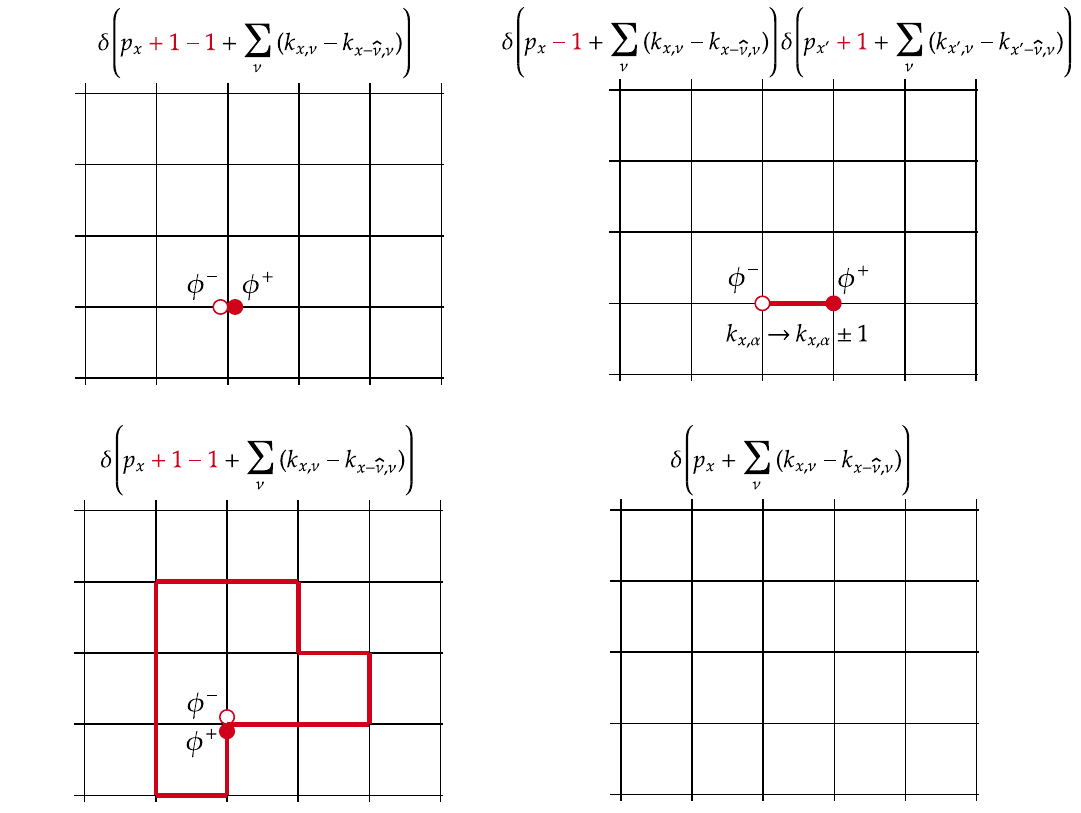}
    \caption{An illustration of the worm update in the $k$-sector. The first step is the insertion of the external source-sink pair $\phi^+\phi^-$. Then the head $\phi^+$ is moved to a neighboring site $x'=x+\hat{\alpha}$ and $k_{x,\alpha}$ is shifted by $\pm1$ to compensate. This is continued until the head and tail are again on the same site when they can be removed. Alternatively, if $\phi^-$ had been chosen as the head, the process would be the same except the signs would be reversed. } 
    \label{fig:worm_update}
\end{figure*}

The worm algorithm was first introduced in the context of spin models~\cite{Prokofev:2001ddj} by Prokof'ev and Svistunov as a substitute for cluster algorithms \cite{Swendsen:1987ce} to combat critical slowing-down. Beyond applications in spin models~\cite{Wolff:2008km}, it has been used, for instance, in the study of $\phi^4$ theories \cite{Gattringer:2012df,Gattringer:2012ap,Rindlisbacher:2016zht}, Gauge--Higgs systems \cite{Schmidt:2012uy,DelgadoMercado:2012tte}, $\On{N}$ models \cite{Endres:2006xu,Bruckmann:2015sua,Rindlisbacher:2015xku,Rindlisbacher:2017ysn,Katz:2016azl}, and the $\mathbb{C}P(N-1)$ model \cite{Bruckmann:2015sua,Rindlisbacher:2016cpj}. We use the algorithm from  \cite{Rindlisbacher:2015xku,Rindlisbacher:2016zht,Rindlisbacher:2017ysn}. As the basic algorithm is described extensively in the references, we only cover its general structure here on a moderately detailed level. We also introduce aspects of the algorithm that have not been described elsewhere yet. 

The basic idea behind a worm algorithm is to temporarily allow for defects in the constraints. First, a pair of defects is placed on the same site, so that they compensate for each other. Then, one defect is moved around the lattice from site to site using importance sampling with local updates. The dual variables on the links along which the defect moves are also updated. When the defects are once again on the same site, the algorithm can be terminated by removing them. This entire process constitutes a single worm update. With them, it is possible to update, with good acceptance rates, dual variables along a non-local closed path so that the constraints remain satisfied.

For the charged sector, the insertion of a defect pair, that compensate for each other, means placing a source and sink pair $\phi^+_x\phi^-_x$ into the system. In our formalism, doing this at some random site $x$ corresponds to adding a $\pm 1$ pair into the matching delta function constraint (\ref{eq:deltafunctionconstraint}),
\begin{equation}
    \delta(p_x\textcolor{red}{+\,1-1}+\sum_\mu(k_{x,\mu}-k_{x-\hat{\mu},\mu})) \ .
\end{equation}
As the total change in the expression is zero, the constraint is still satisfied.

The worm update proceeds by proposing to move the $\phi^+$, which we call the head of the worm, to a random neighboring site $x'=x+\hat{\alpha}$, while the tail of the worm, $\phi^-$ remains at $x$. In terms of the delta functions, the $+1$ is moved from the delta function at $x$ to the one at $x'$
\begin{multline}
    \delta(p_x\textcolor{red}{-1}+\sum_\nu(k_{x,\nu}-k_{x-\hat{\nu},\nu}))\\
    \cdot\delta(p_{x'}\textcolor{red}{+1}+\sum_\nu(k_{x',\nu}-k_{x'-\hat{\nu},\nu})) \ .
\end{multline}
Now, the expressions equal $+1$ and $-1$ for $x'$ and $x$, respectively, and the constraints are violated at both sites. However, if simultaneously to the head moving from $x$ to $x'$, we shift the flux variable on the corresponding link
by plus or minus one, $k_{x,\alpha}\rightarrow k_{x,\alpha}\pm1$, depending on whether $\alpha$ is a positive or negative direction, the expressions in the delta functions remain zero. 

The head is moved around the lattice in this way, updating the $k$-variables along its path, until head and tail are once again located on the same site $x$. Then a removal of the source and sink pair can be attempted, which corresponds to a removal of the $\pm1$ pair from the constraint at $x$. The entire worm update is illustrated in figure~\ref{fig:worm_update}. The source $\phi^-$ can also be chosen as the head. Then the $-1$ will move between the constraints and the $k$ variables will be shifted accordingly with opposite signs.

The update works similarly for the $\chi^{\of{i}}$ sector. The insertion of the $\phi^i_x\phi^i_x$ pair corresponds to adding two $+1$s into the evenness constraint (\ref{eq:evennessconstraint}) at $x$,
\begin{equation}
\sum\limits_{\nu}\of{\chi^{\of{i}}_{x,\nu}+\chi^{\of{i}}_{x-\hat{\nu},\nu}}+n^{\of{i}}_x\textcolor{red}{+\,1+1} \ .
\end{equation}
Either one of them can then be considered as worm head and moved around the lattice. As the constraints simply require the expressions $\sum_{\nu}\ssof{\chi^{\of{i}}_{x,\nu}+\chi^{\of{i}}_{x-\smash{\hat{\nu},\nu}}}+n^{\of{i}}_x$ to be even, the $\chi^{\of{i}}$ can be shifted by either $+1$ or $-1$ to keep the constraints satisfied regardless of the direction. However, since $\chi^{\of{i}}_{x,\nu}\ge0$, updates that would make a $\chi^{\of{i}}$-variable negative are rejected.

When the worm head and tail are present in the system during a worm update, the algorithm samples the relative weights of configurations corresponding to two-point partition functions. If, \eg, the head is located at a site $x$ and consists of a sink of type $i$, and the tail is located at $x_0$ and consists of a source of type $j$, the configuration belongs to the ensemble of the two-point partition function $Z^{i\,j}_2(x,x_0)$. Once the worm update has finished and the source and sink have been removed, the system is again in a configuration from the ensemble of the partition function $Z$. Therefore, when measuring observables on the ensemble of $Z$, they have to be evaluated between worm updates. The worm updates themselves can be understood as sampling two point functions, 
\begin{equation}
\avof{\phi^i\of{x_0}\phi^j\of{x}}=Z^{i\,j}_2(x_0,x)/Z\ .
\end{equation}

Our worm algorithm explicitly samples the transitions between $Z$ and $Z_2^{i,j}\of{x_0,x_0}/V_{\Lambda}$, $i,j\in\cof{1,\ldots,N}$, where $V_{\Lambda}$ is the lattice size, \ie, the total number of sites. It can therefore be interpreted as sampling a generalized partition function,
\begin{equation}
    Z_\text{gen}=Z+\frac{1}{V_{\Lambda}}\sum^N_{i,j=1}\sum_{x,y} c_{ij}(y-x)\,Z^{i\,j}_2(x,y)\ , \label{eq:zgen}
\end{equation}
and therefore we call it the \emph{generalized worm} algorithm. The weights $c_{ij}(y-x)$ can be optimized to achieve volume-filling worm updates in which all head-tail-separations, $y-x$, are sampled with equal probabilities, as explained in more detail in Appendix~\ref{app:GenPartFandWL}. 

\begin{figure}[!ht]
    \centering
    \includegraphics[width=0.48\textwidth]{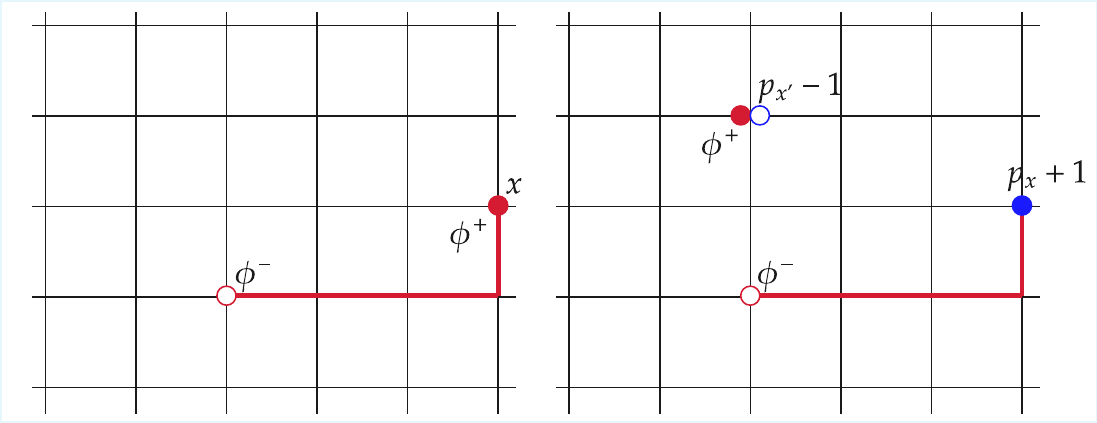}
    \caption{A diagram of a disconnected worm move. By shifting $p_x$ and $p_{x'}$ in the $k$-sector or $n^{\of{i}}_x$ and $n^{\of{i}}_{x'}$ in the $\chi^i$ sector, the worm head can be removed and reinserted into any site on the lattice. The variable shifted at the new site $x'$ can also be of a different type than the one at $x$, which allows for changing the head type.} 
    \label{fig:worm_update_disc}
\end{figure}

If the simulation is performed with non-zero source terms, then instead of changing the flux variable between $x$ and $x'$ to compensate for moving the worm head from $x$ to $x'$, the monomer variables $p$ and/or $n^{\of{i}}$ on these two sites could be changed to avoid defects. In this case, it is also possible to move the head to a new site $x'$ which is not a neighbor of the current head position $x$. So, in these \emph{disconnected} worm moves, the head can be moved to any site on the lattice. Even the head type can be changed by shifting one type of monomer variable at $x$ and a different type at $x'$. As the $p$ and $n^{\of{i}}$ variables stem from a power series expansion of the source term exponentials~\eqref{eq:s+pow.ser.}, \eqref{eq:smpow.ser.}, and \eqref{eq:sipow.ser.}, the disconnected worm moves are only available in sectors where the source has a non-zero value. A disconnected move is visualized in figure \ref{fig:worm_update_disc}.

With non-zero source terms, it is also possible to collect measurements for estimating condensate expectation values $\avof{\phi^{i}\of{x}}=Z^i_1\of{x}/Z$ during the worm updates. The weights of configurations contributing to the one-point partition functions $Z^i_1(x_0)$ and $Z^j_1(x)$ can be obtained from configurations contributing to the two-point partition function $Z_{2}^{i\,j}\of{x_0,x}$ by reweighting away the source $\phi^i\of{x_0}$ or the sink $\phi^j\of{x}$. This means computing the reweighting factors for replacing either the source at $x_0$ by an appropriate shift in the monomer variable $p_{x_0}$ or $n^{\of{i}}_{x_0}$, or, equally, for replacing the sink at $x$ by a shift in $p_{x}$ or $n^{\of{j}}_{x}$.

Our worm algorithm works for both the linear and non-linear $\On{N}$ model. The magnitudes $S_x=|\phi_x|$ of the fields are present only in the weights $W_{\lambda,N}(n)$ \eqref{eq:weightflambda} in the dual variable representation of the partition function \eqref{eq:onfluxreppartf}. As mentioned above, the non-linear case corresponds to taking the limit $\lambda\rightarrow\infty$. In this limit, the only non-vanishing contribution from \eqref{eq:weightflambda} is at $S=1$ where the integrand is independent of $n$. So, the weight function $W_{\lambda,N}(n)$ can simply be replaced by a constant in \eqref{eq:onweightfunc0} to obtain the partition function of the non-linear $\On{N}$ sigma model. We have set this constant to 1.

\section{Simulation algorithms for entanglement entropy in lattice $O(N)$ models} \label{sect:latt_EE}

To compute the derivative of entanglement entropy \eqref{eq:seelsderiv} on the lattice, both the $\ell$ and the $r$ derivative must be approximated as finite differences. By approximating $\partial_r$ with a discrete forward derivative, followed by setting $r\to1$, one approximates $\SEE$ by the second R\'{e}nyi entropy \eqref{eq:renyientropy}. By also approximating $\partial_{\ell}$ with a finite difference, $\dSEE$ becomes
\begin{align} \label{eq:dSEE_Latt}
       \left. \frac{\partial \SEE(\ell')}{\partial \ell'} \right\vert_{\ell'=\ell+1/2}\approx\left. \frac{\partial H_2(\ell')}{\partial \ell'} \right\vert_{\ell'=\ell+1/2} \nonumber\\
       \approx\tilde{\Omega}\of{\ell,2}-\tilde{\Omega}\of{\ell+1,2} \ .
\end{align}
Now, if the number of spatial dimensions is larger than one, the change from $\ell$ to $\ell+1$, present in \eqref{eq:dSEE_Latt}, is a highly non-local change. This means that the ensemble of configurations corresponding to the partition function $\tilde{Z}(\ell,2)$ will in general have very little overlap with the ensemble of configurations corresponding to $\tilde{Z}(\ell+1,2)$, causing a so-called \emph{overlap problem} when trying to sample the free energy difference in \eqref{eq:dSEE_Latt} with Monte Carlo. A way around this, introduced in \cite{Rindlisbacher:2022bhe,Jokela:2023rba}, is to make the update algorithm move back and forth along a sequence of updates, where the boundary between $A$ and $B$ is deformed piece by piece in a specific order to interpolate between $\tilde{Z}(\ell,2)$ and $\tilde{Z}(\ell+1,2)$. The order is chosen so that the free energy differences between subsequent steps are minimized. During the interpolation, histograms $h_i$ are accumulated for how often the system is in each boundary state $i$. The derivative of $\SEE$ can then be evaluated from these histograms as
\begin{equation} \label{eq:dSEE_hist}
       \left. \frac{\partial \SEE(\ell')}{\partial \ell'} \right\vert_{\ell'=\ell+1/2}\approx \log(h_{N})-\log(h_0) \ ,
\end{equation}
where $h_0$ corresponds to the number of configurations encountered with the boundary at $\ell$ and $h_N$ to the number of configurations with the boundary at $\ell+1$. The histograms are stored and reset periodically, so that statistical uncertainties in \eqref{eq:dSEE_hist} can be estimated with jack-knife resampling.

There are alternative approaches to addressing the overlap problem between ensembles at different $\ell$. For example, the interpolation between $\tilde{Z}(\ell,2)$ and $\tilde{Z}(\ell+1,2)$ can also be obtained with an interpolating action, \begin{equation}
\tilde{S}_{\alpha}=\of{1-\alpha}\,\tilde{S}\of{\ell,2}+\alpha\,\tilde{S}\of{\ell+1,2}\ ,\ \alpha\in\fof{0,1}\ .
\end{equation}
This is used to either integrate numerically the expectation value $\ssavof{\partial_\alpha \tilde{S}_{\alpha}}_{\tilde{Z}\of{\alpha}}$, evaluated on the ensembles of the interpolating partition function,
\begin{equation}
\tilde{Z}\of{\alpha}=\int\DD{\phi}\,\e^{-\tilde{S}_{\alpha}\fof{\phi}}
\end{equation}
at sufficiently many $\alpha$~\cite{Buividovich:2008kq,Nakagawa:2009jk,Nakagawa:2010kjk,Itou:2015cyu,Rabenstein:2018bri}, or to perform non-equilibrium Monte Carlo measurements~\cite{Alba:2016bcp,Bulgarelli:2023ofi,Bulgarelli:2024onj}. The difference of free energies in \eqref{eq:dSEE_Latt} can also be computed with normalizing flows~\cite{Bulgarelli:2024yrz, Bulgarelli:2025ewp}. 
It is also worth mentioning that the overlap problem can often be avoided if the entanglement measure is not evaluated on the full vacuum ensemble but on a part of it that has overlap with some operator, \ie, if the entanglement measure is evaluated on an observable in the presence of a static quark-antiquark pair~\cite{Amorosso:2024leg,Amorosso:2024glf,Amorosso:2026mdo,Amorosso:2026zkj} for example. Tensor-network methods provide another independent route to entanglement measures~\cite{Coser:2013qda,Yang:2015rra,Bazavov:2017hzi,Cataldi:2023xki,Hayazaki:2025srr} by being able to access density matrices directly.

\begin{figure}[!htb]
    \centering
    \includegraphics[width=0.5\textwidth]{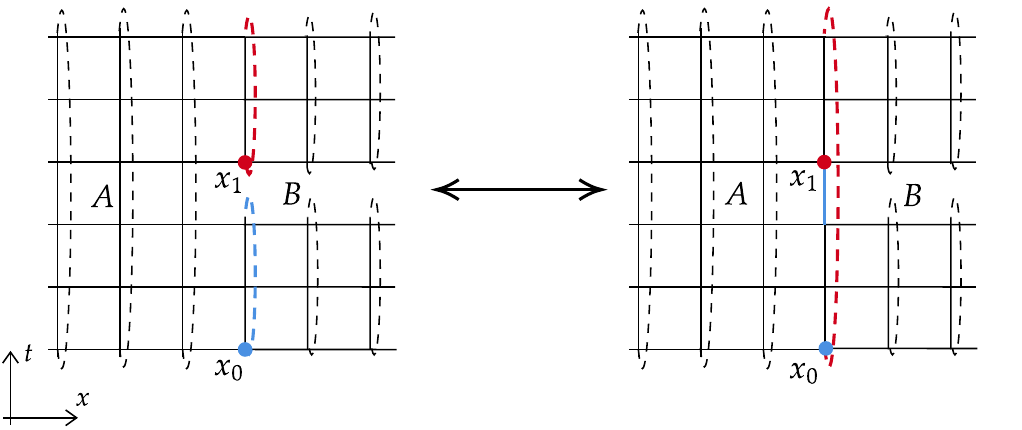}
    \caption{The figure shows how the endpoints of temporal links ending at the bottom corners of the replicas $x_{0,1}$ are swapped when the boundary conditions over the corresponding spatial point are changed. This can lead to violations of the delta function and evenness constraints in the partition function \eqref{eq:onfluxreppartf}.}
    \label{fig:BC_updates}
\end{figure}

The constraints in the dual variable partition function \eqref{eq:onfluxreppartf} also affect the evaluation of $\dSEE$. As the boundary is deformed and the temporal boundary conditions over a spatial site $\bar{x}$ change, the end points $x_{j}=\ssof{\bar{x},j\,N_t-1}$, $j=0,\ldots,r-1$ of $r$ temporal links are exchanged (see the illustration for $r=2$ replicas in figure~\ref{fig:BC_updates}).
Consequently, the temporal flux variables that connect to the sites $x_0,\ldots,x_{r-1}$ from the negative time direction change. For the $l$ variables, this simply means that a corresponding acceptance test is required. With the variables $k$ and $\chi^{\of{i}}$, $i=3,\ldots,N$, things are not as straightforward. If the $k$ or $\chi^{\of{i}}$ fluxes that arrive at the sites $x_0,\ldots,x_{r-1}$ from the negative time direction change value or parity respectively during this change of temporal boundary conditions, defects will be produced, \ie, the delta function or evenness constraints on the sites $x_0,\ldots,x_{r-1}$ will be violated. Therefore, the $k$ and $\chi^{\of{i}}$ sectors need to be operated on before or after the boundary conditions are changed. This can be done by either manipulating the values of the exchanged flux variables so that no defects are caused by the boundary flip or by removing the defects after the flip with worm updates. We have developed update algorithms for both approaches.

Next, we will describe the general structure of both of these updates. Technical details on detailed balance and acceptance probabilities can be found in appendix~\ref{app:probabilities}. The two updates produce fully compatible results and a comparison of their computational costs is provided in appendix~\ref{app:updates}.

\begin{figure*}[ht]
    \centering
    \includegraphics[width=0.8\textwidth]{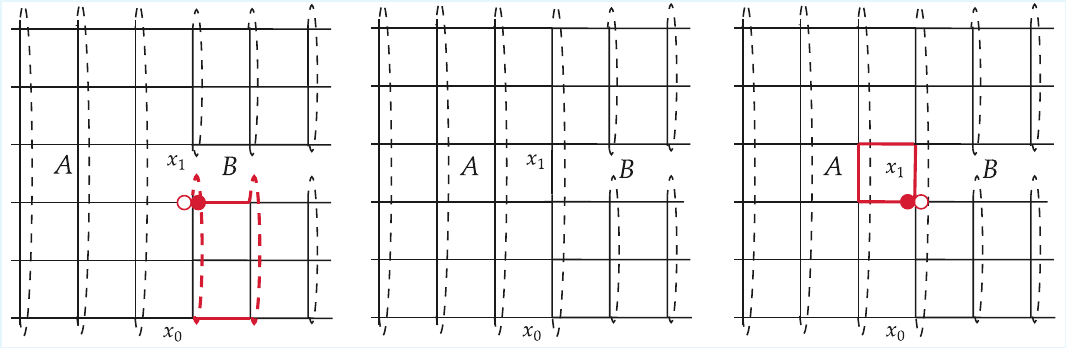}
    \caption{A visualization of the plaquette boundary update. Worm updates constrained to a temporal plaquette are performed to change the value of the problematic flux variables, so that the difference between the variables, whose endpoints are exchanged, is zero for $k$ and zero modulo two for $\chi^{\of{i}}$. This will allow the boundary conditions to be changed without issues. After the flip, the same plaquettes are performed in reverse to restore the differences to the values they had originally in order to maintain detailed balance.}
    \label{fig:plaq}
    \includegraphics[width=0.8\textwidth]{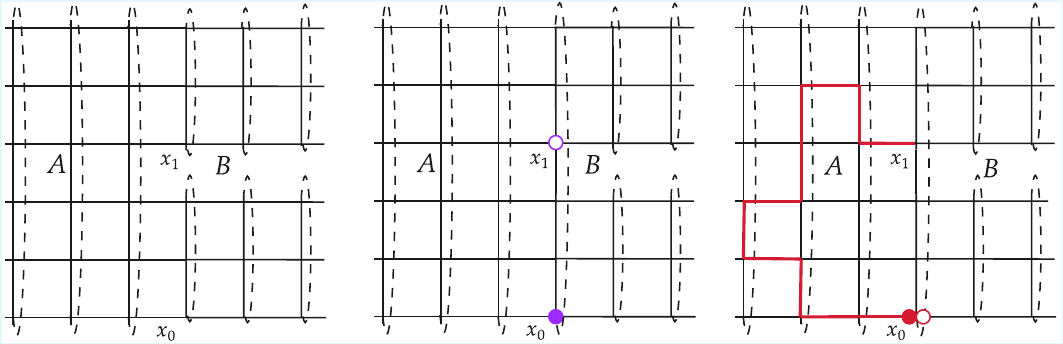}
    \caption{A visualization of the boundary worm update. A pair of defects introduced by the change of temporal boundary conditions consists of a defect (filled purple disk) on site $x_0$ and an anti-defect (open purple disk) on site $x_1$ (middle panel). A worm (red) can move the defect from site $x_0$ around until it eventually reaches the site $x_1$ (right panel), after which the defects can be removed.}
    \label{fig:BC_worm}
\end{figure*}

\subsection{Boundary plaquette update}\label{ssec:plaquetteupdate}

We can modify the values of the problematic flux variables $k_j=k_{x_j-\smash{\hat{d},d}}$ and $\chi^{\smash{\ssof{i}}}_j=\chi^{\smash{\ssof{i}}}_{x_j-\smash{\hat{d},d}}$, $j=0,\ldots,r-1$ directly with a sequence of worm update, where the head is constrained to move along a temporal plaquette that contains the temporal link between $x_j$ and $x_j-\hat{d}$ (see figure~\ref{fig:plaq}). Whenever such a \emph{plaquette worm} update succeeds, \ie, the head of the worm moves around the whole plaquette, the problematic flux variable changes value by $\pm1$. For the $k$ variables, the sign of the change is controlled by the charge the worm head carries and whether the link is gone through in the positive or negative direction. The whole update routine performs the required plaquette worm updates to make 
\begin{equation}
\Delta k_j=k_{\sigma\of{j}}-k_{j}=0
\end{equation}
and
\begin{equation}
\Delta \chi_j^{\smash{\of{i}}}=\chi^{\smash{\of{i}}}_{\sigma\of{j}}-\chi^{\smash{\of{i}}}_{j}=0 \ (\text{mod} \ 2)\ ,
\end{equation}
for all $j=0,\ldots,r-1$, where $k_{\sigma\of{j}}$ and $\chi^{\smash{\of{i}}}_{\sigma\of{j}}$ are the link variables that would respectively replace $k_{j}$ and $\chi^{\smash{\of{i}}}_{j}$ at the site $x_j$ after the boundary condition update. Then, the update of the boundary conditions, \ie, the exchange of the link variables, can be carried out without violating any constraints. After changing the boundary conditions, the program executes the same plaquette worms in reverse but with the new boundary conditions to restore the $\Delta k_{j}$ and the parities of the $\Delta \chi_j^{\smash{\ssof{i}}}$ to their original values. This guaranties detailed balance, as the system is put in a state where the reverse boundary update, if selected, would trigger the same set of plaquette worm updates.

In the plaquette update routine, it is first randomly decided whether a spatial site is added or removed from $A$. Then, the program calculates how many plaquette worms are needed in each sector to make $\Delta k_j$ and $\Delta \chi_j^{\of{i}} \bmod{2}$ zero for this spatial site. At this stage, it is also randomly decided for each plaquette worm whether it is performed around $x_j$ or $x_{\sigma\of{j}}$. The head and tail pairs are inserted one step in the negative time direction from these sites. A head can go through the spacetime sites of a plaquette in two orientations. The orientations are dictated at the start of the update routine according to the directions in which the problematic flux variables need to be shifted to make $\Delta k_j$ or $\Delta \chi_j^{\smash{\ssof{i}}} \bmod{2}$ smaller. The spatial orientations of the plaquettes need to be such that all points in a plaquette have the same temporal boundary condition, \ie, they are in the same region of the system (either $A$ or $B$). The spatial directions are chosen randomly from the ones satisfying this constraint during the insertions of the head and tail pairs at the starts of the plaquette worms.  With the necessary plaquette worms, the program constructs a chain of updates. The first part of the chain contains the plaquette worms that will make the differences $\Delta k_j$, $\Delta \chi_j^{\smash{\ssof{i}}} \bmod{2}$ zero. Then follows an update which consists of an acceptance test for the $\Delta l_j$, $\Delta \chi_j^{\smash{\ssof{i}}}$.\footnote{An acceptance test is needed for the $\chi^{\smash{\ssof{i}}}$ since the plaquette worms make these differences only zero modulo 2.} If it is accepted, the temporal boundary conditions of the spatial site are changed. After the change of boundary conditions, the second part of the chain is performed, which is equivalent to the first sequence of plaquette worms but in reverse order and with the new boundary conditions. The program goes through the chain in a snake-like manner~\cite{deForcrand:2000qn}, choosing at each point randomly whether to attempt to move forward or backward along the chain of updates.

\subsection{Boundary worm update}\label{ssec:boundarywormupdate}

The second approach is based on the fact that the defects produced during a change of temporal boundary conditions over a spatial site $\bar{x}$ always appear in pairs of opposite types, located on different replica corners $x_0,\ldots,x_{r-1}$. We can therefore use worm updates to move defects form one replica corner to another to bring defects of opposite types, \ie, a defect and its \emph{anti-defect}, together, which then compensate each other (cf. figure~\ref{fig:BC_worm}). 

When adding a site to region $A$ the program first changes the boundary conditions. This causes the defect-anti-defect pairs to appear on different replica corner pairs. Let us assume that one such defect-anti-defect pair appears at the sites $x_j$ and $x_{\sigma\of{j}}$. The program then performs special worm updates, where an appropriate worm head and the corresponding tail are inserted at the sites $x_j$ and $x_{\sigma\of{j}}$, to compensate the defects. The head is then moved around the system until it reaches the location of the tail, after which they can be removed. This is repeated till all defect-anti-defect pairs have been removed from the system. To ensure detailed balance, the update can also terminate when the head and tail are at the different replica corners. When this happens, the update is considered to have failed, as the defects reappear. A list of worm updates required to remove all defects is determined at the beginning of the boundary update, as well as a list of required on-site acceptance tests, due to the $l$ and $\chi^{\smash{\ssof{i}}}$ variables being exchanged. This list is usually not unique, since multiple defect-anti-defect pairs of the same type are produced during a boundary condition update, so that different worm head-tail-pair combinations are possible for their removal. The shuffled list of all required updates forms a chain which is again gone through with a snake algorithm. The boundary condition update is successful if the snake algorithm terminates after traversing the whole chain of updates. If the snake algorithm terminates after having returned to the starting point, the update has failed and the old boundary conditions are restored.

When a site is removed from $A$, the procedure is essentially the same but in reverse, meaning that for the worm updates, head and tail start now on the same replica corner, say $x_\sigma\of{j}$, after which the head aims to reach the replica corner $x_j$. If the worm ends when the head is located on $x_j$ the worm update was successful, while if the worm ends when the head is located again on the same site as the tail, the worm updated has failed. In this way, successful worm updates introduce defect pairs in such a way that they get removed when the boundary conditions are switched. 

Detailed balance requires that for each possible boundary update, the corresponding inverse update has a finite selection probability. To ensure that this is the case, the boundary worm updates are not allowed to change the flux variables $k_j=k_{x_j-\smash{\hat{d},d}}$ and $\chi^{\smash{\ssof{i}}}_j=\chi^{\smash{\ssof{i}}}_{x_j-\smash{\hat{d},d}}$, $j=0,\ldots,r-1$, \ie, the boundary worms are forbidden from moving along the temporal links that connect from the negative time direction to the sites $x_j$, $j=0,\ldots,r-1$. In this way, the defect pattern, \ie, the number and location of the different defect types, encountered when preparing a reverse boundary update, will be the same as the one encountered by the forward boundary update.  

During the boundary worm update both connected and disconnected moves are possible. However, since we are not interested in sampling the full two point function during the boundary worm update, head changing disconnected moves are not performed; these are not required for ergodicity and would just increase the update time since the worm would move in a larger configuration space, making it harder to find the desired end state.

\section{On selection of simulation parameters}\label{sec:select}

The purpose of the simulations carried out in this work is to verify the relations between entanglement and R\'enyi entropies, and thermodynamic quantities, as discussed in Sec.~\ref{sect:EEthermo}. These relations were derived without referring to a specific type of system and should therefore hold for any sufficiently well behaved one. In particular, the identities should hold in lattice field theories at finite lattice spacing. 
To verify the relations, we will therefore not aim for simulation parameter sets that would enable a continuum extrapolation of the results, but simply select simulation parameters for which the required measurements can be carried out at moderate computational cost.    

\begin{figure}[ht]
    \centering
    \includegraphics[width=0.75\linewidth]{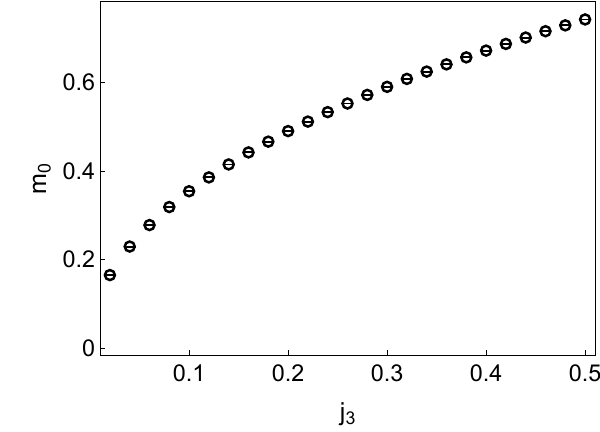}\\[3pt]
    \includegraphics[width=0.75\linewidth]{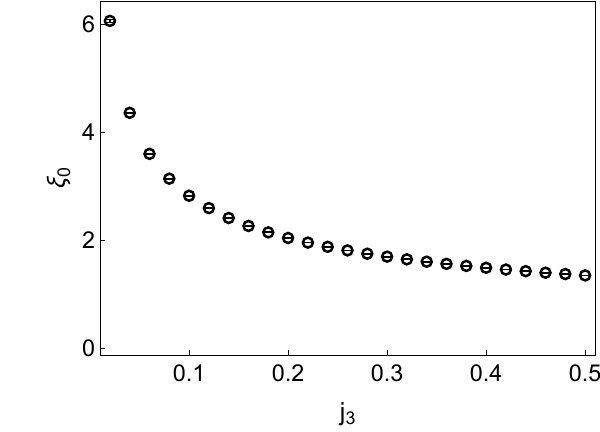}
    \caption{Mass $m_0$ (top) and corresponding correlation length $\xi_0=1/m_0$ (bottom) of the three degenerate pseudo-Goldstone modes in the three-dimensional non-linear lattice $\On{4}$ model at $\kappa=1.2$ and $\mu=0$, as a function of the source $j_3$. The data was obtained on lattices of size $N_s^2\,N_t=18^3$.}
    \label{fig:goldstonemassvsj3}
\end{figure}

\begin{figure}[ht]
    \centering
    \includegraphics[width=0.75\linewidth]{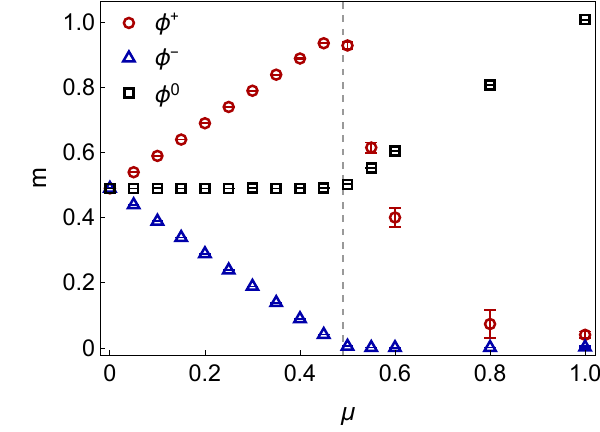}\\[3pt]
    \includegraphics[width=0.75\linewidth]{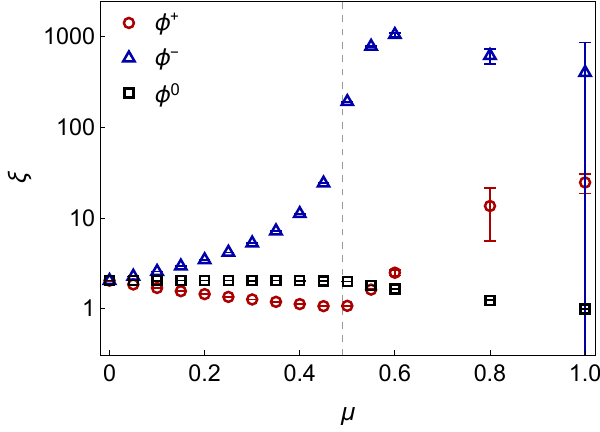}
    \caption{Mass spectrum (top) and corresponding correlation lengths (bottom) of the three-dimensional non-linear lattice $\On{4}$ model at $\kappa=1.2$, $j_3=0.2$ as functions of $\mu$ on a lattice of size $N_s^2\,N_t=18^3$. Note that the $\phi^+$ mass is only accurately determined up to the critical $\mu\approx 0.5$ (dashed, vertical line), since at finite density, $\phi^+$ has overlap with the vacuum and is no longer a well-defined particle state. Also the correlation length measurement for $\phi^{-}$ is no longer accurate after growing beyond the IR cutoff scale at $\mu\approx 0.5$ (dashed, vertical line).}
    \label{fig:massspectrum}
\end{figure}

As such, we choose to perform our simulations in $d=\of{2+1}=3$ dimensions. We set the hopping parameter to $\kappa=1.2$, which places the system well inside the phase where the global $\On{4}$ symmetry is spontaneously broken to $\On{3}$. In this regime, the radial mode is heavy and effectively decouples,\footnote{Since we are simulating the \emph{non-linear} $\On{4}$ model, the radial mode is actually infinitely heavy.} leaving three Goldstone modes, $\sscof{\phi^+, \phi^-, \phi^0}$, as the relevant low-energy degrees of freedom. Since relations \eqref{eq:felatlargelapprox} and \eqref{eq:thermsdensmaxrel} require a finite correlation length, we also introduce a source $j_3=0.2$, which gives the Goldstone modes a mass $m_0\approx 0.5$ at $\mu=0$. The dependencies of $m_0$ and the corresponding correlation length $\xi_0=1/m_0$ on the value of $j_3$ are illustrated in figure~\ref{fig:goldstonemassvsj3}. In 4D, our $\On{4}$ action would be equivalent to the leading principal chiral action of two-flavor isospin QCD~\cite{Son:2000by,Rindlisbacher:2017ysn}, with the decoupled radial mode being the $\sigma$ particle, the $\sscof{\phi^+, \phi^-, \phi^0}$ being the three pions, and with $j_3$ playing the role of a degenerate quark mass term. In this context, figure~\ref{fig:goldstonemassvsj3} is the $\of{2+1}$-dimensional lattice analog of the well-known Gell-Mann--Oakes--Renner 
relation, $m_{\pi}\propto \sqrt{m_u+m_d}$.  

For $\mu>0$ the masses of $\sscof{\phi^+, \phi^-, \phi^0}$ split, as shown in the upper panel of figure~\ref{fig:massspectrum}, and the mass of the lightest mode, $\phi^{-}$, behaves as
\begin{equation}
    m^{-}\of{\mu}=\ucases{m_0-\mu\quad\text{if}\quad\mu<m_0\\0\quad\text{otherwise}}\ .
\end{equation}
The longest correlation length in the system is, therefore, given by
\begin{equation}
\xi_{\text{max}}\of{\mu}=1/m^{-}\of{\mu}\label{eq:maxcorrlen}
\end{equation}
and diverges as $\mu$ approaches the critical value $\mu_c=m_0\approx 0.5$ (cf. figure~\ref{fig:massspectrum}, bottom). This critical value corresponds to the finite density phase transition.

The data in the figures~\ref{fig:goldstonemassvsj3} and \ref{fig:massspectrum} is produced from lattice simulations of systems without replicas. These unreplicated simulations are also used to estimate the right-hand side of \eqref{eq:maxwell_rel_hr}. They are carried out on lattices of size $N_t\,N_s^{d-1}$, with $N_s=18$ and $N_t=5,\ldots,20$. Each of these simulations consists of $2\cdot 10^7$-$10^8$ full updates.

The systems with $r=2$ replicas used to compute $\dHn{2}$ were simulated on lattices of size $r\,N_t\,V$ with spatial lattice volume $V=N_x\,V_{\perp}$, where $V_{\perp}=N_s^{d-2}$ with $d=3$ and $N_s=12$, and $N_x=36$. Furthermore, $N_t=5,\ldots,10$ to capture several different temperatures. The width $\ell$ of the entangling region is measured along the $x$-direction. To obtain the general structure of $\dHn{2}$ as a function of $\ell$, we measured $\dHn{2}$ for a few values of $N_t$ and $\mu$ on several intervals $\fof{\ell,\ell+1}$ for $1\le\ell\le17$. To obtain results in the regime where $\xi\ll\ell\ll N_x$ is satisfied, necessary for the relation \eqref{eq:eederiveqtherms} to hold, we focused our simulations on the interval $\ell\in\fof{17,18}$ with significantly more statistics and a much wider range of $N_t$ and $\mu$ values.

Simulations with both boundary update algorithms discussed in sections~\ref{ssec:plaquetteupdate} and \ref{ssec:boundarywormupdate} were used to obtain the results. As the algorithms produce fully consistent results, we do not explicitly mention which algorithm was used to produce which data. The amount of updates $N_u$ performed to obtain the $\dHn{2}$ data can be seen in table~\ref{tab:EEstats}. Of the updates, around $1/4$ are boundary updates and the rest are generalized worm updates. The first two rows of the table correspond to the simulations performed on multiple different $\ell$ intervals. The last two rows correspond to the large $\ell$ simulations, where $\ell\in\fof{17,18}$.
\begin{table}[ht]
\centering
 \begin{tabular}{|C{3cm}|C{2cm}|} 
    \hline
    &$N_u$ \\
    \hline
    $\ell$-scan, $N_t=5$& $7.5\times10^8$ \\
    $\ell$-scan, $N_t=8$ & $4.5\times10^8$ \\
    large $\ell$, $N_t<10$& $1.5\times10^9$ \\
    large $\ell$, $N_t=10$ & $1.2\times10^9$\\ 
    \hline
 \end{tabular}
 \caption{Table listing the number of updates $N_u$ performed in each simulation for the $\dHn{2}$ measurements. Boundary updates amount to about 1/4 of the total number of updates; the remaining 3/4 are generalized worm updates, \ie, sweeps. The first two rows are for simulations performed with several different $\ell$ intervals. The last two are for the large $\ell$ case, where $\ell$ is interpolated between $\ell=17$ and $18$.}
 \label{tab:EEstats}
\end{table}

\begin{figure*}[ht]
    \centering
\includegraphics[width=0.99\linewidth]{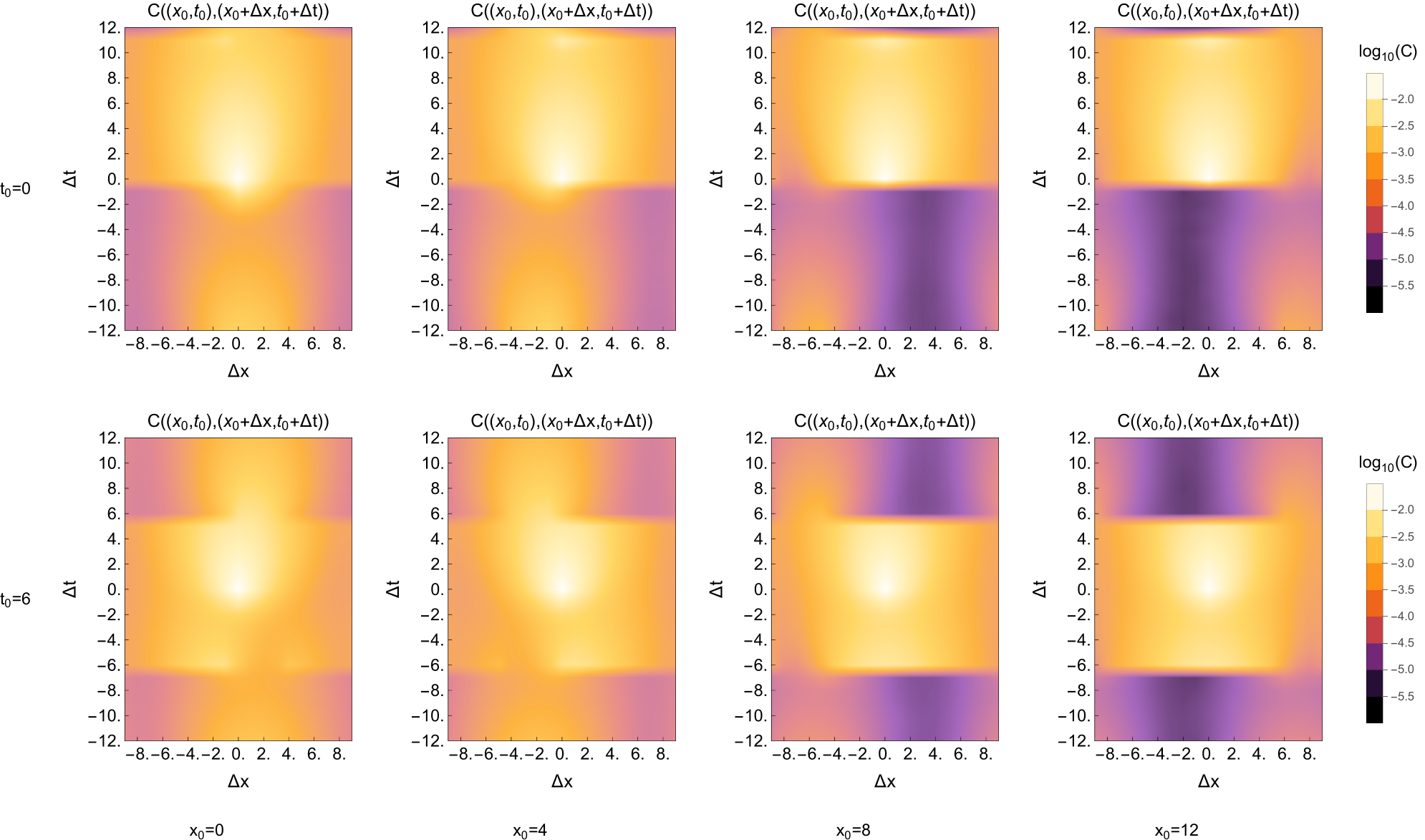}
\caption{In a replicated system as described by \eqref{eq:latlandauferep}, translation invariance is broken in $x=x^1$ and $t=x^d$ direction. The figure shows the projection of the charged two-point function $C\of{\of{x_0,t_0},\of{x_0+\Delta x,t_0+\Delta t}}=\avof{\phi^-\of{x_0,t_0}\,\phi^+\of{x_0+\Delta x,t_0+\Delta t}}$ to the $\Delta x$-$\Delta t$-plane for a non-linear $\On{N}$ model as described by partition function~\eqref{eq:onfluxreppartf} with $N=4$, $\kappa=1.2$, $j_3=0.5$, and $\mu=0.3$ on a $\of{2+1}$-dimensional $r=2$ replica lattice of size $N_x\times N_y\times r\cdot N_t=18\times12\times 2\cdot 12$ and with entangling region width set to $\ell=4$. The different panels correspond to different choices of $\of{x_0,t_0}$. If $\of{x_0,t_0}$ is within region $B$ (over which the temporal boundary conditions separate the two replica), propagation from one replica to the other is highly suppressed. This is due to the fact that the disconnected piece of the shown two-point function is zero for the given parameter values and the only way to propagate in region $B$ from one replica to the other is therefore via a detour through region $A$. The non-zero chemical potential manifests in the plots through a higher probability to propagate forward in Euclidean time than backward. The two-point function was measured during optimized worm update, as discussed in Appendix~\ref{app:GenPartFandWL}.}
\label{fig:correlationfunctiondensplot}
\end{figure*}

\section{Results}\label{sec:results}

Figure~\ref{fig:correlationfunctiondensplot} shows the charged two-point correlation function $\avof{\phi^-\of{x_0}\phi^+\of{x}}$ for the non-linear, $\of{2+1}$-dimensional lattice $\On{4}$ model, simulated in a $r=2$-replica system of size $N_x\times N_y\times r\cdot N_t=18\times12\times 2\cdot 12$, using the simulation parameter values $\kappa=1.2$, $j_3=0.5$, and $\mu=0.3$, and entangling region width $\ell=4$. 
Since translation invariance is broken in $x=x^1$ and $t=x^d$ direction in this system, the two-point function is shown in the $\of{x,t}$-plane as function of the source sink separations $\Delta x=x-x_0$ and $\Delta t=t-t_0$ for different source locations $\of{x_0,t_0}$, 
\begin{multline}
C\of{\of{x_0,t_0},\of{x_0+\Delta x,t_0+\Delta t}}=\\
\avof{\phi^-\of{x_0,t_0}\,\phi^+\of{x_0+\Delta x,t_0+\Delta t}}\ .
\end{multline}
The two-point function behaves as one would expect in this setup: if $\of{x_0,t_0}$ is within region $B$ (over which the temporal boundary conditions separate the two replica), propagation from one replica to the other is highly suppressed. This is due to the fact that the disconnected piece of the shown two-point function is zero for the given parameter values and the only way to propagate in region $B$ from one replica to the other is therefore via a detour through region $A$. The non-zero chemical potential manifests itself in the plots through a higher probability to propagate forward in Euclidean time than backward.

\begin{figure}[htb]
    \centering
    \includegraphics[width=0.435\textwidth]{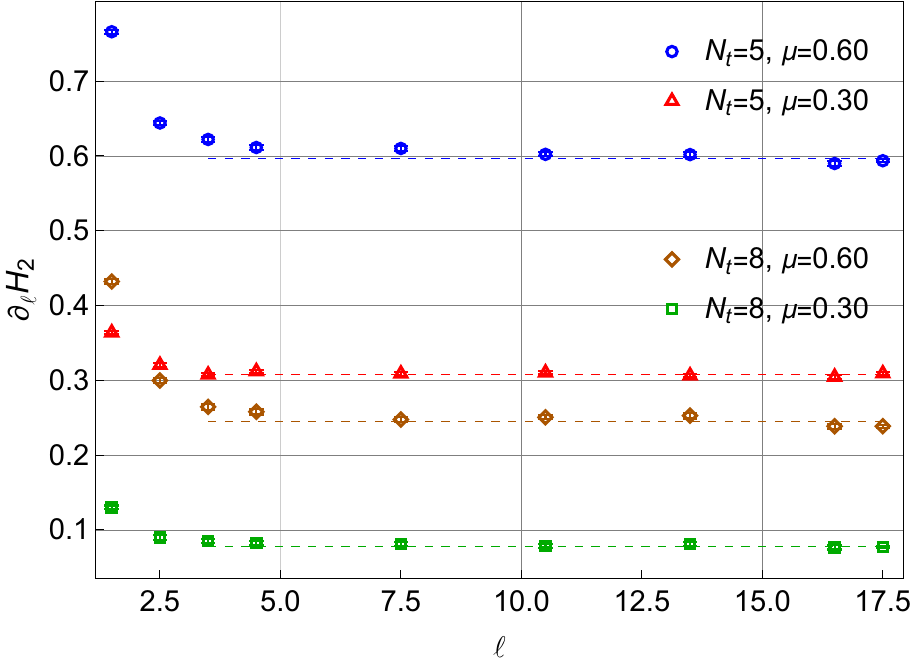}
    \caption{$\dHn{2}$ of the three-dimensional non-linear $\On{4}$ model, with the simulation parameters described in section \ref{sec:select}, as a function of $\ell$ for $N_t\in\{5,8\}$ and $\mu\in\{0.30,0.60\}$. $\dHn{2}$ quickly reaches a $N_t$ and $\mu$ dependent plateau value marked in the plot with dashed lines. As argued in section~\ref{sect:EEthermo}, this value corresponds, up to a factor of $1/V_{\perp}$, to the $r=2$ step scaling approximation \eqref{eq:stepscaleentropydens} of the thermal entropy density, \ie, $\dHn{2}/V_{\perp} = s_2\of{T\of{N_t},\mu}$, if $\xi\ll\ell\ll N_x$.}
    \label{fig:lscan}
\end{figure}

The $\ell$ dependence of $\dHn{2}$ for the three-dimensional non-linear $\On{4}$ model, with the simulation parameters described in section \ref{sec:select}, can be seen in figure~\ref{fig:lscan} which shows $\dHn{2}$ as a function of $\ell$ for $N_t\in\{5,8\}$ and $\mu\in\{0.30,0.60\}$. In all four cases, $\dHn{2}$ initially decreases rapidly as function of increasing $\ell$ until around $\ell\geq 5$, where it reaches a $N_t$ and $\mu$ dependent plateau value. According to relation~\eqref{eq:hromegaexpr} from section~\ref{sect:EEthermo}, this value corresponds, up to a factor of $1/V_{\perp}$, to the step scaling approximation $s_2\of{T\of{N_t},\mu}$ (cf. \eqref{eq:stepscaleentropydens}) of the thermal entropy density $s\of{T,\mu}$, if $\xi\ll\ell\ll N_x$.

\begin{figure}[!htb]
    \centering
    \includegraphics[scale=0.375]{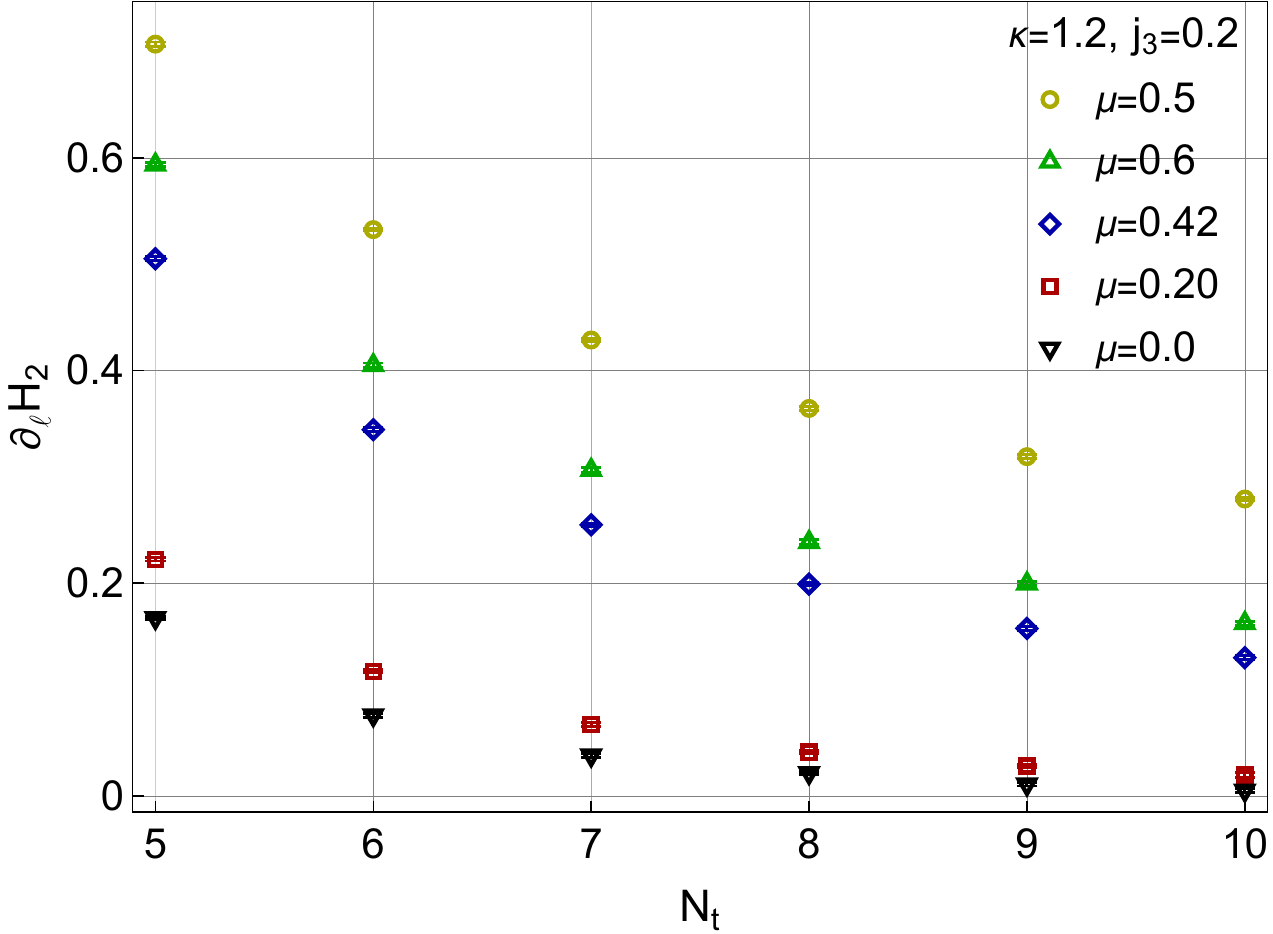}
    \caption{$\dHn{2}$ of the three-dimensional non-linear $\On{4}$ model as a function of $N_t$ for $\mu\in\{0.0;0.20;0.42;0.5;0.6\}$ at $\ell=\ell^*:=17.5$. According to \eqref{eq:hromegaexpr}, this corresponds to $V_{\perp}\,s_2\of{T\of{N_t},\mu}$ for the $N_t$ and $\mu$ values for which the condition $\xi\ll\ell\ll N_x$ is satisfied. $\dHn{2}$ decreases with increasing $N_t$ for all $\mu$ values.}
    \label{fig:dSEE_nt}
\end{figure}

\begin{figure}[!htb]
    \centering
    \includegraphics[scale=0.375]{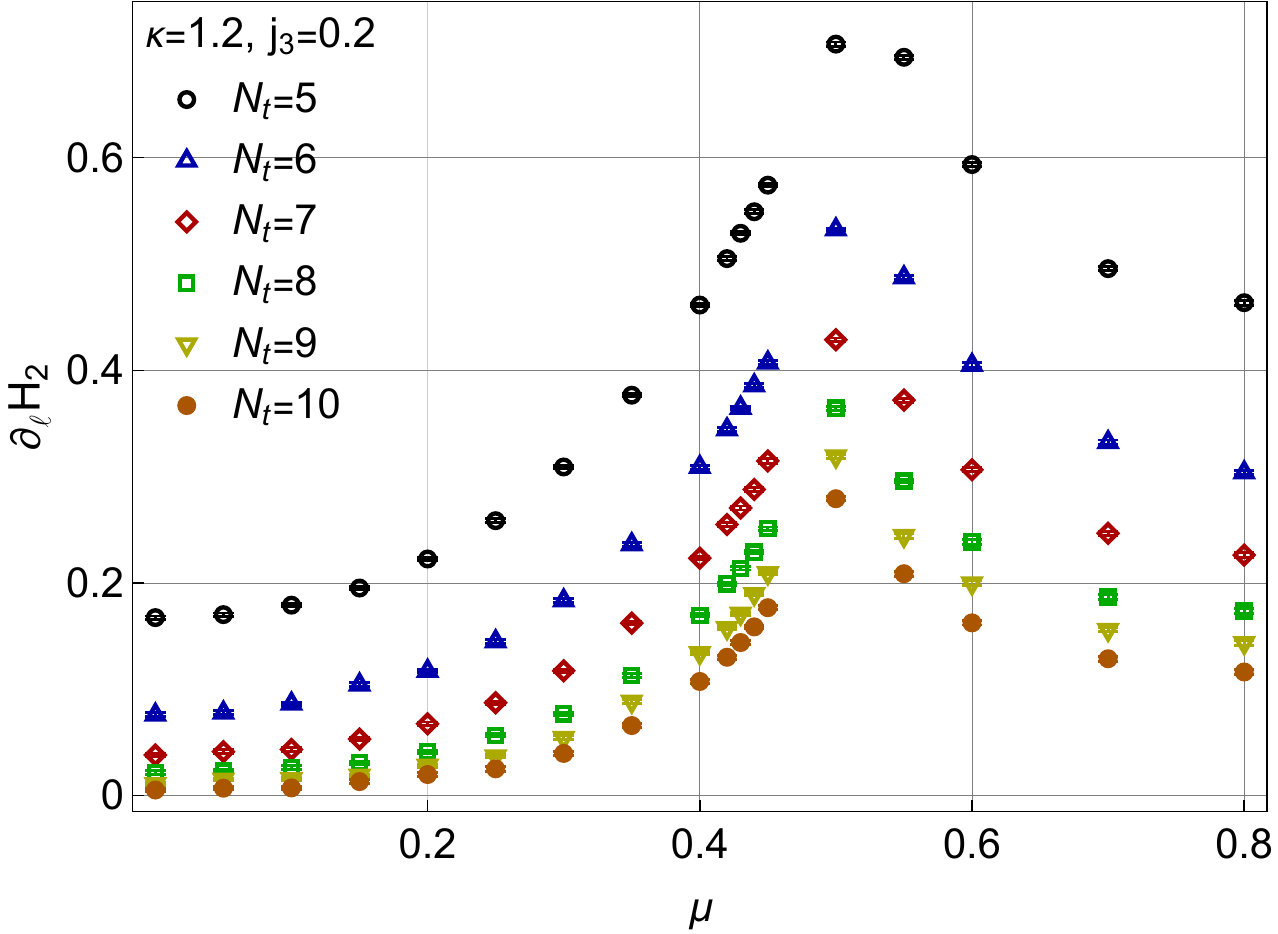}
    \caption{ $\dHn{2}$ of the three-dimensional non-linear $\On{4}$ model as a function of $\mu$ for $N_t\in\{5,6,7,8,9,10\}$ at $\ell=\ell^*=17.5$. According to \eqref{eq:hromegaexpr}, this corresponds to $V_{\perp}\,s_2\of{T\of{N_t},\mu}$ for the $N_t$ and $\mu$ values for which the condition $\xi\ll\ell\ll N_x$ is satisfied. The phase transition around $\mu_c\approx0.5$ is clearly visible in the plot.}
    \label{fig:dSEE_mu}
\end{figure}

The  $N_t$ and $\mu$ dependency of $\dHn{2}$ is presented in figures~\ref{fig:dSEE_nt} and~\ref{fig:dSEE_mu}, where the $\dHn{2}$ data at the largest $\ell$-value $\ell=\ell^*:=17.5$, obtained by performing the entangling region boundary deformation to interpolate between entangling region widths $\ell=17$ and $\ell=18$, is plotted as a function of $N_t$ for $\mu\in\{0.0,0.2,0.42,0.5,0.6\}$ in figure~\ref{fig:dSEE_nt} and as a function of $\mu$ for $N_t\in\{5,6,7,8,9,10\}$ in figure~\ref{fig:dSEE_mu}. As mentioned earlier, for the $N_t$ and $\mu$ values for which $\xi\ll\ell\ll N_x$ with $\ell=\ell^*$, these $\dHn{2}\vert_{\ell=\ell^*}$ values correspond to $V_{\perp}\,s_2\of{T\of{N_t},\mu}$. It can be seen that as $N_t$ increases, $\dHn{2}\vert_{\ell=\ell^*}$ decreases. The relation between it and $\mu$ is more complicated. $\dHn{2}\vert_{\ell=\ell^*}$ increases with $\mu$ until the critical value $\mu=\mu_c\approx0.5$ is reached, after which $\dHn{2}\vert_{\ell=\ell^*}$ decreases if $\mu$ is increased further. The phase transition around $\mu_c$ is clearly visible in the $\mu$-dependency of $\dHn{2}\vert_{\ell=\ell^*}$, which highlights the ability of entanglement measures to probe phase transitions.

To test the relations \eqref{eq:thermsdensmaxrel} and \eqref{eq:maxwell_rel_hr}, the charge density has to be defined on the lattice in terms of the dual variables. We will explicitly distinguish between the charge density of the standard grand canonical partition function $Z(\beta,V,\mu)$
\begin{equation} \label{eq:lattice_n}
    n=\frac{1}{\vlatt}\frac{\partial \log Z}{\partial\mu}=\frac{\langle \sum_xk_{x,d}\rangle}{\vlatt} \ 
\end{equation}
and of the one with the $r$-replica geometry $\tilde{Z}(\ell,\beta,V,\mu,r)$
\begin{equation} \label{eq:lattice_n_2}
    \nrep\of{\ell,r}=\frac{1}{r\,N_t V}\frac{\partial \log \tilde{Z}(\ell,r)}{\partial\mu}=\frac{\langle \sum_xk_{x,d}\rangle_{r}}{r\,N_t V} \ .
\end{equation}
In the previous equations, the $\beta,V,\mu$ arguments were suppressed and the expectation value is evaluated in the replicated configuration $\langle \mathcal{O}\rangle_{r}\equiv \int \mathcal{D}[\phi]\, \mathcal{O}\fof{\phi}\,e^{-\tilde{S}_{r,\ell}[\phi]}/\tilde Z\of{\ell,r}$ in the latter case. The required sum of $k$-variables can be evaluated at almost no computational cost.

The charge density $\nrep$ can be measured in a straight forward manner for any fixed value of $\ell$ . Doing so for multiple $\ell$ values, one can take a numerical derivative of $\nrep$ with respect to $\ell$ to compute the mixed second derivative of $\log\tilde{Z}$ with respect to $\ell$ and $\mu$, \emph{i.e.},
\begin{equation}
    \frac{1}{\vrlatt}\frac{\partial^2(\log \tilde{Z})}{\partial\ell\,\partial\mu} = \partial_{\ell}\nrep\ ,
\end{equation} 
where we also set $r=2$. Since $\dHn{2}=-\partial_\ell\log \tilde{Z}$, it then follows that
\begin{equation}
    \frac{\partial^2H_2}{\partial\mu\,\partial\ell}=-\vrlatt \frac{\partial \nrep}{\partial\ell}\ . \label{eq:d2F}
\end{equation}
The left hand side of \eqref{eq:d2F} can also be evaluated directly by measuring $\dHn{2}$ for a set of $\mu$ values and then taking the numerical derivative with respect to $\mu$. 
By comparing the results obtained with these two evaluation methods for $\partial_{\ell}\partial_{\mu}\log\tilde{Z}$, we can check the correctness of our boundary deformation algorithm. 
\begin{figure}[!htb]
    \centering
    \includegraphics[scale=0.375]{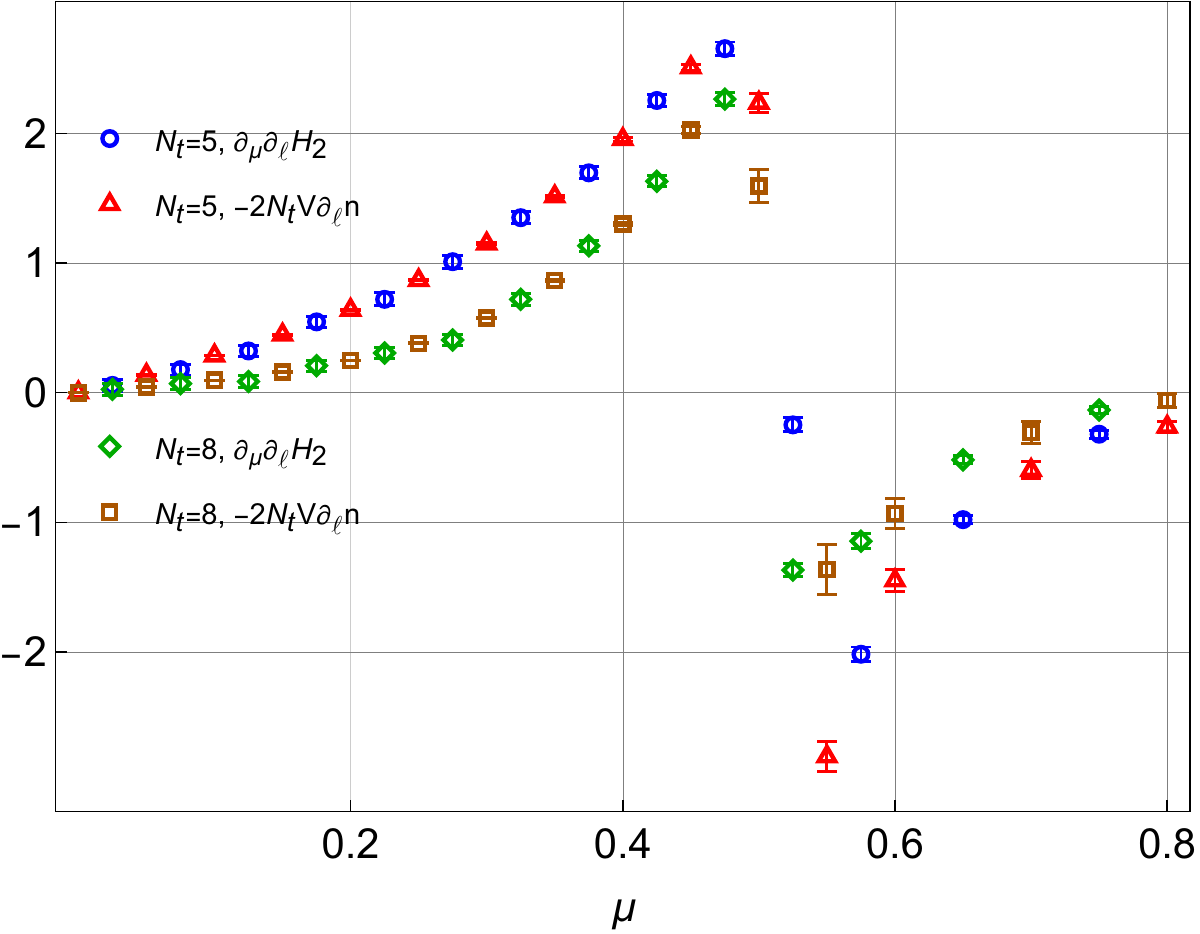}
    \caption{Comparison of results for $-\frac{\partial^2(\log \tilde{Z})}{\partial\mu\partial_\ell}$ as a function of $\mu$ in the three-dimensional non-linear $\On{4}$ model as computed from $\partial_\mu\dHn{2}$ and $-\vrlatt \partial_\ell \nrep$ for two $N_t$ values. Both cases show a clear agreement between the two evaluations.}
    \label{fig:derivatives}
\end{figure}
Such a comparison is shown in figure~\ref{fig:derivatives} which depicts both sides of the relation \eqref{eq:d2F} as functions of $\mu$ for $N_t\in\{5,8\}$. Only two $N_t$ values are shown to keep the plot clear and readable. For both $N_t$ values, the results obtained by taking the $\mu$-derivative of $\dHn{2}$ and the ones obtained by taking the $\ell$ derivative of $\nrep$ show excellent agreement, which is also the case for all other $N_t$ and $\mu$ values we simulated, as well as arbitrary values of the other simulation parameters, $\kappa$, $j$, $\ldots$. This indicates that our algorithm is working properly.

Interestingly, based on \eqref{eq:d2F}, one could evaluate $\dHn{2}$ by integrating $\partial_\ell\tilde{n}$ measurements with respect to $\mu$. This may be beneficial in systems, where the boundary deformation is costly, as $\partial_\ell\tilde{n}$ can be evaluated with static values of $\ell$. 

Recall that on the lattice, we use the approximation $\dSEE\approx\partial_{\ell}H_2$. Therefore, to verify the validity of the relation \eqref{eq:thermsdensmaxrel}, we instead have to focus on \eqref{eq:maxwell_rel_hr} with $r=2$, \ie, on
\begin{equation}
\frac{1}{V_{\perp}}\frac{\partial^2 H_2}{\partial\mu\,\partial\ell}\overset{{}_*}{=}-2\,N_t\,\of{n\of{2\,N_t,\mu}-n\of{N_t,\mu}}\ ,\label{eq:maxwell_rel_h2}
\end{equation}
where the asterisk ($*$) refers to the requirement $\xi_{\text{max}}\ll\ell,N_x/2,N_s$ for the equality to hold.

Both sides of the relation \eqref{eq:maxwell_rel_h2} can be computed from lattice simulations. The left-hand side can be obtained from simulations with $r=2$ as explained previously, while the right-hand side follows from measurements of the charge density \eqref{eq:lattice_n} in pairs of unreplicated systems of temporal extent $N_t$ and $2\,N_t$, respectively.

\begin{figure}[!htb]
    \centering
    \includegraphics[scale=0.375]{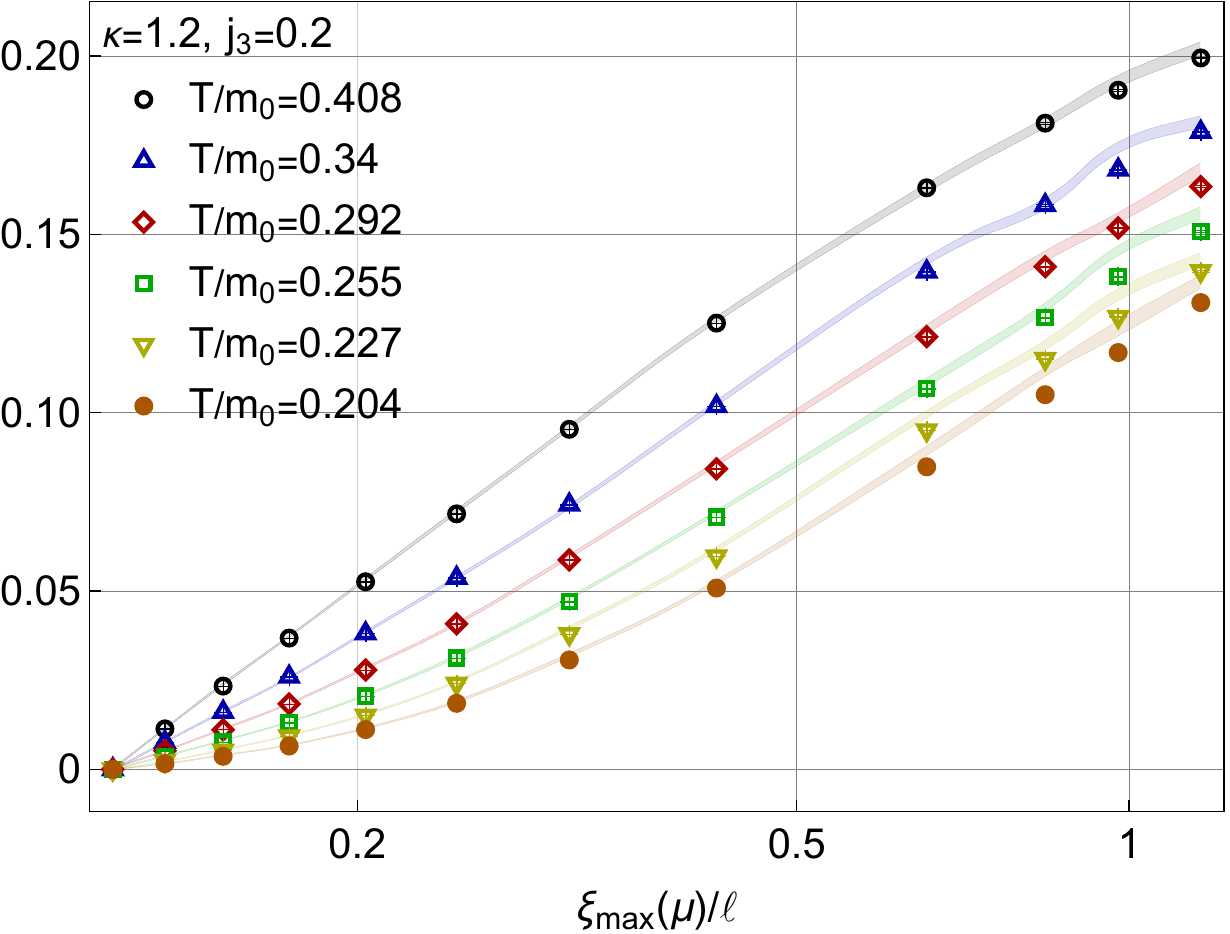}
    \caption{Comparison of the two sides of \eqref{eq:maxwell_rel_h2} in the three-dimensional non-linear $\On{4}$ model. The bands show $V_\perp^{-1}\partial_\mu\partial_{\ell}H_2$ (evaluated via $\partial_\ell \nrep$), while the points denote $-2\,N_t\,\of{n\of{2\,N_t,\mu}-n\of{N_t,\mu}}$, plotted as functions of $\xi_{\text{max}}\of{\mu}/\ell$, with $\ell=\ell^*=17.5$. 
    Agreement persists up to $\xi_{\text{max}}\of{\mu}/\ell\approx 0.5$ at the lowest temperature and almost up to $\xi_{\text{max}}\of{\mu}/\ell\approx 1.0$ for the highest temperature.}
    \label{fig:maxwell_rel_h2}
\end{figure}

A comparison of the two sides of \eqref{eq:maxwell_rel_h2} is given in figure~\ref{fig:maxwell_rel_h2} which shows $V_\perp^{-1}\partial_\mu\partial_{\ell}H_2$ (bands) and $-2\,N_t\,\of{n\of{2\,N_t}-n\of{N_t}}$ (point markers) as functions of $\xi_{\text{max}}(\mu)/\ell$, with $\xi_{\text{max}}\of{\mu}$ given by \eqref{eq:maxcorrlen}. The displayed values of $V_\perp^{-1}\partial_\mu\partial_{\ell} H_2$ have been obtained from measurements of $\partial_\ell\nrep$ through \eqref{eq:d2F}, which yields a slightly cleaner signal for $\mu<\mu_c$ than taking the numerical $\mu$ derivative of $\dHn{2}$. The temperatures are expressed as the dimensionless quantity $T/m_0$ obtained from $1/\of{N_t m_0}$. As can be seen, the two quantities agree well up to $\xi_{\text{max}}/\ell\approx0.5$ for the lowest temperature and almost up to $\xi_{\text{max}}/\ell\approx1.0$ for the highest temperature. This makes sense, since $\xi_{\text{max}}$ refers to the longest correlation length at zero temperature. At finite temperature, the effective correlation length is shortened by thermal effects.

\begin{figure}[!htb]
    \centering
    \includegraphics[scale=0.375]{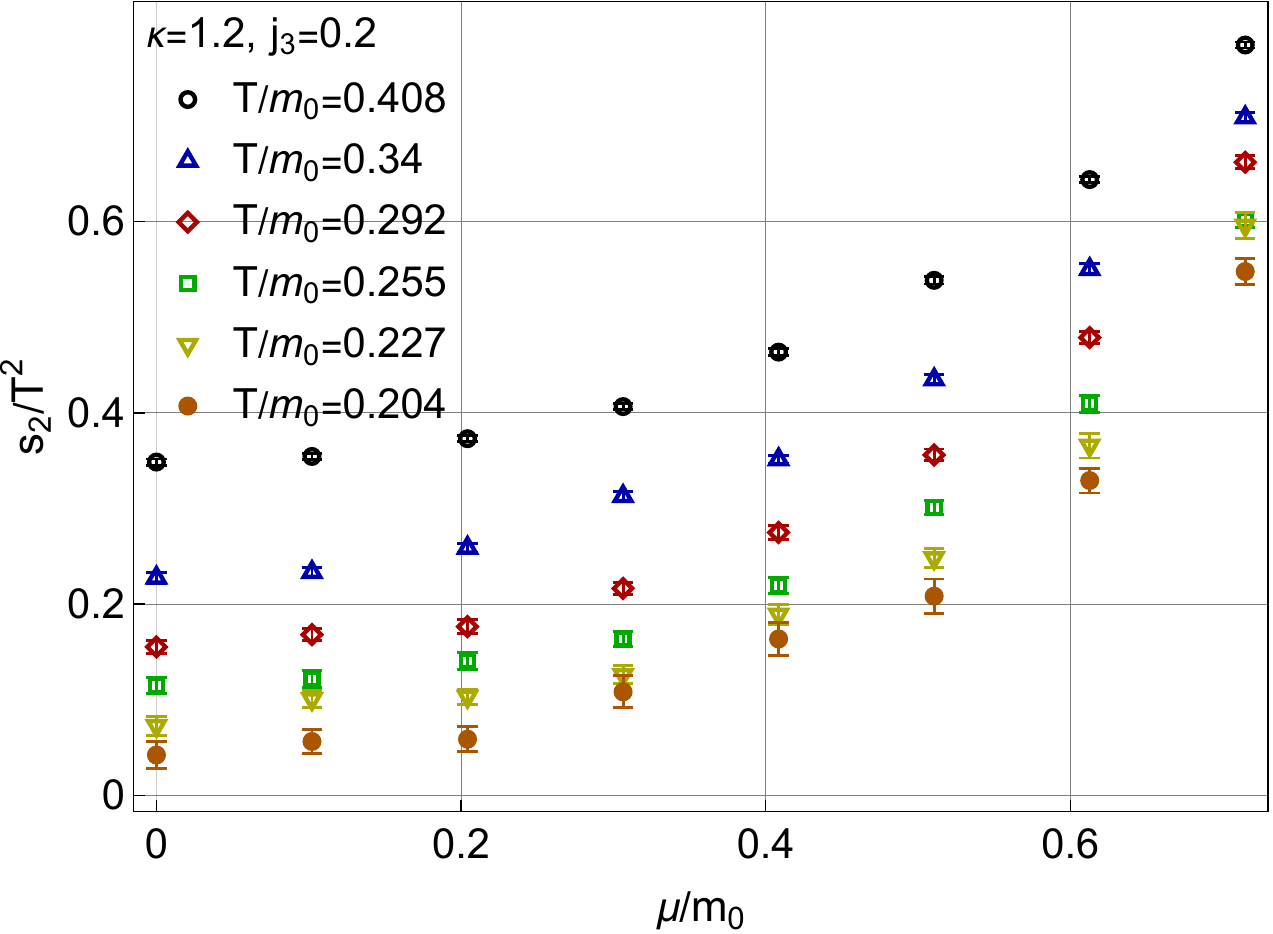}
    \caption{Approximation of the rescaled thermal entropy density $s/T^2$ of the three-dimensional non-linear $\On{4}$ model in terms of the $r=2$ step scaling thermal entropy $s_r$ as a function of $\mu/m_0$. Here $s_r$ has been determined from measurements of $\dHn{2}\vert_{\ell=\ell^*}$ with $\ell^*=17.5$ for $\mu\le 0.4$ by using the relation \eqref{eq:hromegaexpr}.}
    \label{fig:s_mu}
\end{figure}

\begin{figure}[!htb]
    \centering
    \includegraphics[scale=0.375]{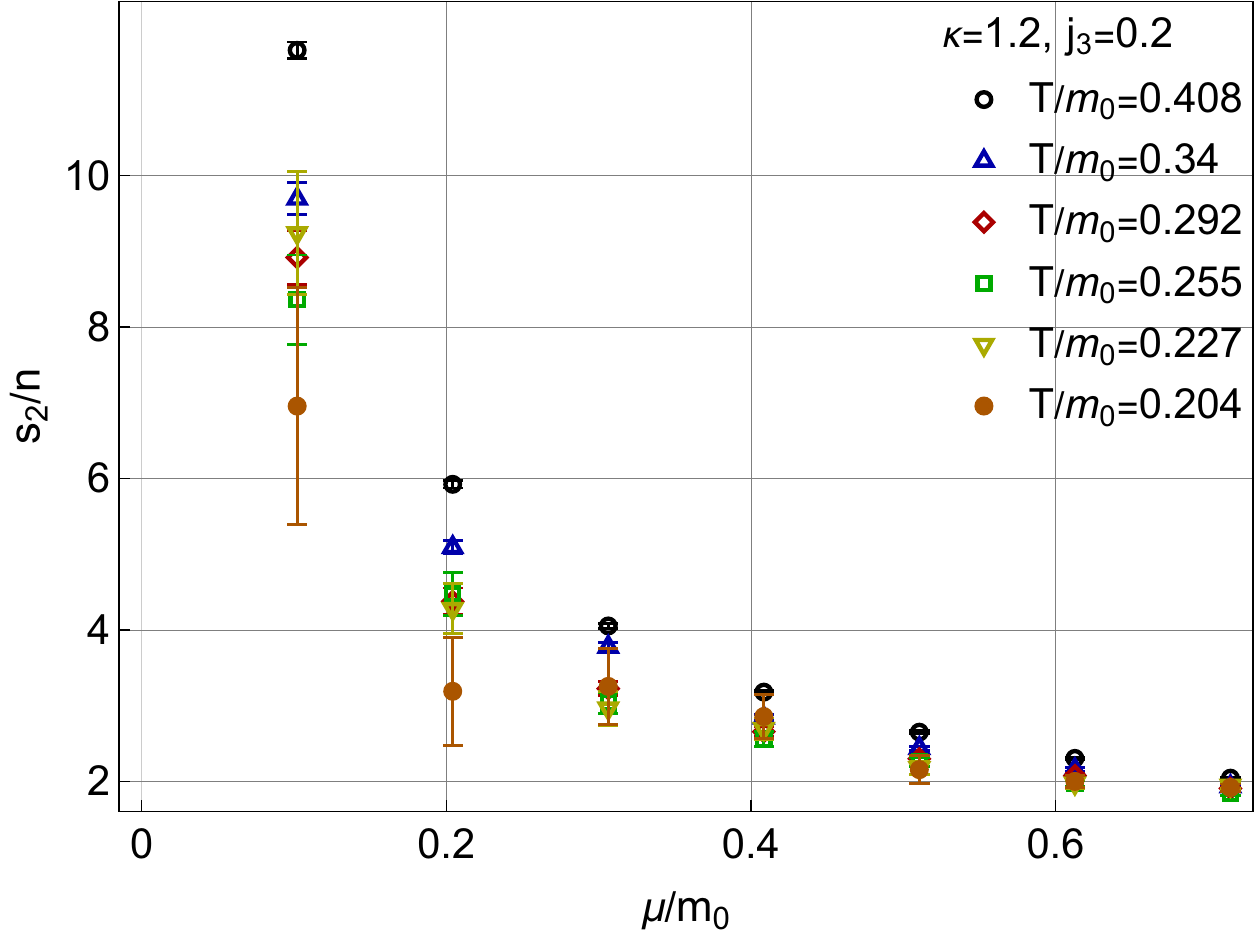}
    \caption{The thermal entropy density of the non-linear three-dimensional $\On{4}$ model scaled with the charge density $s/n$ as a function of $\mu/m_0$, where $s$ has been approximated with the $r=2$ step scaling thermal entropy $s_r$. Here $s_r$ has been determined from measurements of $\dHn{2}\vert_{\ell=\ell^*}$ with $\ell^*=17.5$ for $\mu\le 0.4$ by using the relation \eqref{eq:hromegaexpr}.}
    \label{fig:sn_mu}
\end{figure}

\begin{figure}
    \centering
    \includegraphics[scale=0.375]{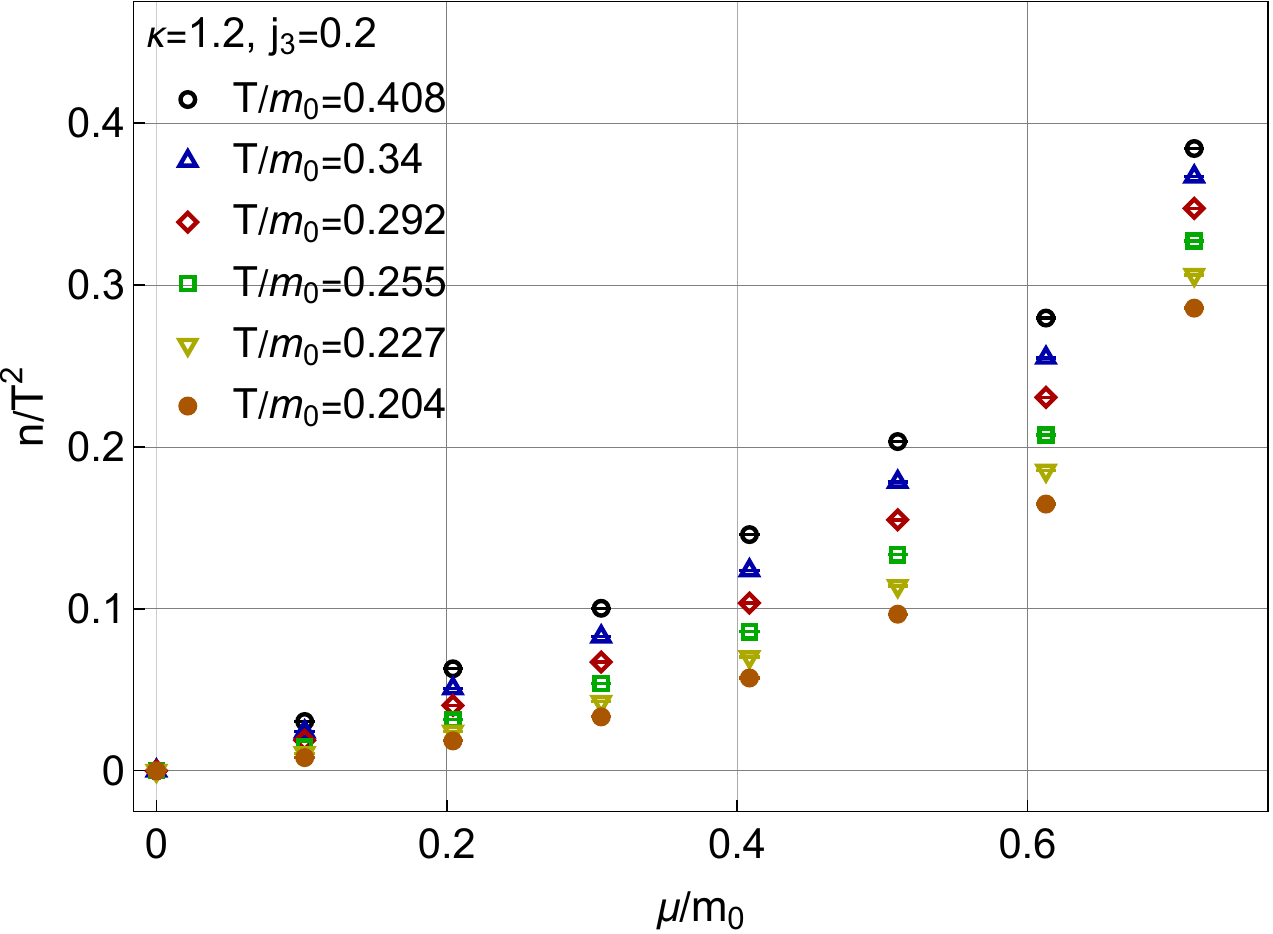}
    \caption{$n/T^2$ as a function of $\mu/m_0$ for the three-dimensional non-linear $\On{4}$ model. The results have been obtained with \eqref{eq:lattice_n} from simulations of unreplicated systems.}
    \label{fig:n_mu}
\end{figure}

Based on these results, we can say that the connection holds at all considered temperatures for at least $\xi_{\text{max}}/\ell\approx0.5$, which corresponds to $\mu\le 0.4$. Therefore, dividing our results of $\dHn{2}$ at $\ell=17.5$ by $V_{\perp}=N_s$, we obtain estimates for the entropy density at these $\mu$ values. The results are presented in figures~\ref{fig:s_mu} and~\ref{fig:sn_mu} which show $s_2/T^2$ and $s_2/n$ as functions of $\mu/m_0$ respectively for the six temperatures considered. Here $s_2$ is the $r=2$ step scaling approximation of the thermal entropy density $s$, as given in~\eqref{eq:stepscaleentropydens}. The values of $n$ have been obtained with \eqref{eq:lattice_n} in simulations of unreplicated systems. These values are visualized in figure~\ref{fig:n_mu} which displays $n/T^2$ as a function of $\mu/m_0$.

\section{Conclusions and outlook}\label{sec:outlook}
In this work, we showed that, for slab-shaped entangling regions, the derivative of the entanglement entropy operator with respect to the entangling region volume approaches the thermal entropy density $s$ in the limit where the linear sizes of the sub-regions are much larger than any correlation length in the system. This relation generalizes to R\'enyi entropies in the sense that, in the same limit, the derivative of the $r$-th R\'enyi entropy with respect to the sub-region volume approaches a step scaling approximation $s_r$ with scaling factor $r$ of the thermal entropy density, as defined in~\eqref{eq:stepscaleentropydens}. These relations are expected to hold for general QFTs. 
We tested this idea with lattice simulations of the three-dimensional $\On{4}$ model at finite density by showing that the derivative, $\dHn{2}$ of the R\'enyi entropy $H_{2}$ with respect to the width $\ell$ of the slab-shaped entangling region, which serves as an approximation for $\dSEE$, satisfies a Maxwell relation of the thermal entropy density if the largest correlation length in the system is smaller than the linear sizes of the entangling region and its complement. 
We identify the parameter space region in which the latter condition is satisfied and use our measurements of $\dHn{2}$ to evaluate the thermal entropy density in this region.

The formalism presented here provides a new way to study nonperturbative QFT thermodynamics with lattice computations supplementing previous approaches \cite{Ding:2015ona,Schmidt:2017bjt}. On that note, thermal entropy density $s$ has been measured before on the lattice for $\SU{N}$ gauge field theories \cite{Panero:2009tv,Borsanyi:2012ve,Giusti:2016iqr,Giusti:2025fxu} and for QCD \cite{Borsanyi:2013bia,HotQCD:2014kol}, but, to our knowledge, not for $\On{N}$ models in general. Our analysis therefore yields, in effect, a nonperturbative determination of the thermal entropy density in the $\On{4}$ model, which can serve as a benchmark for future calculations using conventional thermodynamic methods. Due to this lack of measurements, we tested the proposed framework here through the Maxwell relation rather than by directly comparing $s$ with $\frac{1}{V_\perp}\dSEE$. Extending such direct comparisons to both $\On{N}$ models and gauge theories would be an important next step.

Beyond providing access to thermodynamic quantities, entanglement measures have long been known to be sensitive to phase structure. In particular, derivatives of EE can be used to detect phase transitions and extract critical behavior, although important limitations arise in the large-$N$ limit, where entanglement signatures may become parametrically suppressed~\cite{Nishioka:2006gr,Klebanov:2007ws,Jokela:2020wgs,Jokela:2023lvr,Jain:2025xko}. As such, the critical exponents of the $\On{N}$ universality class \cite{Pelissetto:2000ek} could possibly be evaluated with lattice computations of $\dHn{2}$. The framework developed here also demonstrates that, even away from criticality, EE provides bulk thermodynamic information in a precise and quantitative manner.

An important feature of the relation derived here is that it is most cleanly realized in systems with a finite correlation length, such as confining phases, where the separation between UV boundary contributions and extensive bulk physics is sharp. This regime is notoriously difficult to capture using large-$N$ holographic methods, since confining phases involve parametrically fewer degrees of freedom than their deconfined counterparts. Our results therefore provide a nonperturbative handle on a regime where holographic methods remain subtle, while at the same time offering a natural point of contact with holographic studies of entanglement c-function and related observables in the large-$N$ limit~\cite{Jokela:2025cyz}. Furthermore, the conjectured duality between the three-dimensional critical $\On{N}$ vector model and higher-spin gravity in AdS${}_4$~\cite{Klebanov:2002ja,Giombi:2012ms} presents a fascinating, albeit highly nontrivial, laboratory for exploring finite-$N$ holographic dictionaries. Because the metric in the higher-spin bulk transforms into higher-spin fields under gauge transformations, standard geometric notions break down. Consequently, the bulk gravity dual to EE in this context remains an open theoretical challenge~\cite{deBoer:2013vca,Ammon:2013hba}. Nevertheless, extracting nonperturbative entanglement data directly from the lattice provides the crucial baseline needed to constrain and guide these theoretical developments. While bulk reconstruction from EE has been formally explored elsewhere (see, \eg,~\cite{Bilson:2010ff,Lin:2014hva,Bao:2019bib,Jokela:2020auu,Jokela:2025ime}), extending these inverse problems to the finite-$N$ regime using discrete lattice data is a critical next step. A statistical approach capable of handling the inherently noisy nature of such boundary data to recover the dual metric has been successfully formulated~\cite{Jokela:2020auu,Filev:2026gka,Fan:2025fxt}, and this exercise was rather recently demonstrated using real nonperturbative lattice data for three-dimensional $\SU{2}$ gauge theory~\cite{Jokela:2023rba}. Applying these methods to our $\On{N}$ EE data can thus serve as a useful testbed for precision holography, guiding the recovery of bulk descriptions, and the subsequent prediction of observables like the two-point functions or Wilson loops, even in regimes where classical supergravity approximations break down. 

Several other extensions suggest themselves. The analysis can be repeated in the canonical ensemble, where connections to symmetry-resolved entanglement~\cite{Goldstein:2017bua}, and to its holographic realizations~\cite{Zhao:2020qmn}, become natural. More generally, since the thermal entropy density depends on both temperature and chemical potential, the relations established here imply that EE as defined in \eqref{EE} contains, in principle, the full information needed to reconstruct the equation of state. Exploring this possibility further and extending the analysis to other interacting QFTs are natural directions for future work. 

In this context it is also worth mentioning that when interested in EE as defined in \eqref{EE}, \eg, for holographic applications, the relation between $S_{EE}$ and thermal entropy can also be used in the reverse direction, that is, to obtain $S_{EE}$ from thermal entropy in cases where the latter is easier to compute than $S_{EE}$ itself, \eg, in non-perturbative fermionic theories.

\section*{Acknowledgments}
N.~J. was supported in part by the Research Council of Finland through grant no. 354533 and the Centre of Excellence in Neutron-Star Physics (project 374062). A.~R. was supported by the Finnish Ministry of Education and Culture through the Quantum Doctoral Education Pilot Program (QDOC VN/3137/2024-OKM-4) and the Research Council of Finland through the Finnish Quantum Flagship (project 358878). T.~R. acknowledges support from the European Research Council grant no. 101142449 and Research Council of Finland grant no. 354572. The authors thank CSC - IT Center for Science, Finland, for computational resources.

\appendix
\section{Generalized partition function and Wang-Landau optimization}\label{app:GenPartFandWL}
As stated in section~\ref{subsect:worm}, our worm algorithm samples a generalized partition function $Z_\text{gen}$ consisting of the usual partition function $Z$ and all the two-point partition functions $Z^{i\,j}_2(x,y)$ as follows:
\begin{equation}
    Z_\text{gen}=Z+\frac{1}{V_{\Lambda}}\sum^N_{i,j=1}\sum_{x,y} c_{ij}(y-x)\,Z^{i\,j}_2(x,y)\ , \label{eq:zgen}
\end{equation}
where $V_{\Lambda}$ is the total number of lattice sites. The weights $c_{ij}\ssof{x}=\e^{-g_{ij}\of{x}}$ are optimized at the beginning of a new simulation, using Wang--Landau sampling \cite{Wang:2000fzi} to make the worm updates sample all head-tail-separations equally.

Without optimized weights $c_{ij}(y-x)$, the length of a worm update would depend significantly on the chosen simulation parameters and on which two-point partition function $Z_2^{i\,j}\of{x,y}$ the worm is currently updating. This is evident since $Z_2^{i\,j}\of{x,y} \propto \avof{\phi^{i}\of{x}\phi^{j}\of{y}}$, which means that large head-tail separation $r=\abs{x-y}$ would be suppressed like $\e^{-m\,r}$, where $m$ is the mass of the lightest excitation in the $\of{i,j}$-channel. 

With optimized weights $c_{i\,j}\of{y-x}$, the worm samples all possible head-tail separations equally for any choice of simulation parameters. This is desirable since it means that (a) the number of worm updates performed in a simulation can be used as a simulation parameter independent measure of Monte Carlo time, analogous to the usual definition of a \emph{sweep}; and that (b) the two-point functions sampled during worm updates have equal statistics at all distances, \ie, large distance correlation function measurements do not suffer from exponentially degrading signal-to-noise ratios. 

Two-point correlation functions are measured by accumulating histograms $H$, and $H_{ij}\of{\Delta x}$, which, respectively, keep track of how often the algorithm visits configurations where no worm head-tail-pair is present in the system and how often it visits configurations where a worm is present with a tail of type $i$ and a head of type $j$, separated by $\Delta x$. With the weights $c_{i\,j}\of{\Delta x}$, the two-point correlation functions are obtained from these histograms as
\begin{equation}
\frac{1}{V_{\Lambda}}\sum_{x}\avof{\phi^i\of{x}\phi^j\of{x+\Delta x}}=\frac{H_{ij}\of{\Delta x}}{c_{i\,j}\of{\Delta x}\,H}\ .\label{eq:corrfuncfromwormhist}
\end{equation}

In $r$-replicated systems, used to measure $\dHn{r}\of{\ell}$, translation invariance is broken in the $x=x^1$ and $t=x^d$ directions. In this case, the weights $c_{ij}$ and histograms $H_{ij}$ should be functions not only of the head-tail separation $\Delta x$, but also of the tail location $\of{x_0,t_0}$ in the $\of{x,t}$-plane, and the expression \eqref{eq:corrfuncfromwormhist} had to be adjusted accordingly. However, in our simulations, we do this only for the histogram $H_{ij}$ to resolve the $\of{x_0,t_0}$-dependency of the two-point functions. For the weights $c_{ij}$, we do not add the $\of{x_0,t_0}$-dependency, since it would increase the number weights that need to be optimized by a factor of $\of{N_x\, N_t}$ and render the initial Wang--Landau sampling very time consuming. Fortunately, it turns out that approximately volume filling worms and simulation parameter-independent average worm lengths can also be achieved with weights $c_{ij}\of{\Delta x}$ that depend only on the head-tail separation $\Delta x$, and ignore the broken translation invariance in $x$ and $t$ direction. In our simulations, we therefore optimize and use the ``volume averaged'' definition $c_{ij}\of{\Delta x}$ from \eqref{eq:zgen} for both non-replicated and replicated systems.

\section{Detailed balance and acceptance probabilities in the boundary updates}\label{app:probabilities}
As usual in Monte Carlo simulations, to ensure that our Markov chain truly samples configurations from the desired partition function, we have to ensure that detailed balance is satisfied between each pair of successive configurations $C$ and $C'$.\footnote{This means configurations that can be transformed into each other by a single update. In our case, the individual steps of the worm algorithm and of the boundary updates are all considered single updates.} The general form of the detailed balance equation is 
\begin{equation}
    w(C)P(C\to C')=w(C')P(C'\to C)\ ,
\end{equation}
where $w(C)$ is the weight of the configuration $C$ and $P(C\to C')$ is the transition probability from $C$ to $C'$. Our updates follow the general structure of the \textit{Metropolis--Hastings} algorithm \cite{Hastings:1970aa} with the transition probability factored into a move-choice or selection probability $p$ and an acceptance probability $P_A$
\begin{equation}
    P(C\to C')=p(C\to C')P_A(C\to C')\ .
\end{equation}
Detailed balance can then be satisfied by setting
\begin{equation}
    P_A(C\to C')=\min\left(1,\, \frac{p(C'\to C) w(C')}{p(C\to C') w(C)}\right)\ . \label{eq:MH_acc_prob}
\end{equation}
With symmetric move-choice probabilities, this reduces to the standard \textit{Metropolis} algorithm~\cite{Metropolis:1953am}. In our simulations, however, the number of possible moves changes in different stages of the algorithm. This happens in both the generalized worm update and the plaquette and boundary worm updates. We will go through the move-choice and acceptance probabilities for the boundary updates in this appendix. The probabilities for the generalized worm have been explained in \cite{Rindlisbacher:2017ysn}. As the weights follow rather straightforwardly from the partition function \eqref{eq:onfluxreppartf} and are mostly the same as in the generalized worm case, we will in general not write them out explicitly.

\subsection{Plaquette}
\begin{figure*}[!htb]
    \centering
    \includegraphics[scale=1.25]{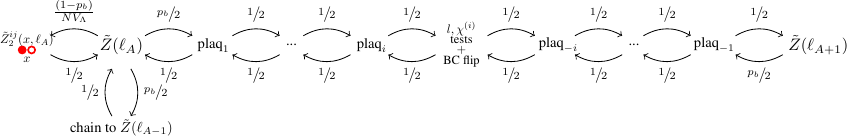}
    \includegraphics[scale=1.25]{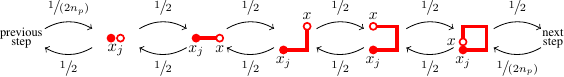}
    \caption{A visualization of the plaquette update chain. The upper graph is for the whole update chain and the lower one is for a single plaquette update. The numbers on the arrows are the selection probabilities of the corresponding updates. The plaquettes with the negative subscript denote the inverse plaquettes performed in the second half of the chain. $\ell_{A+1}$ and $\ell_{A-1}$ are the $\ell$ values corresponding to adding and removing a site from $A$ respectively. The chains have been truncated at the end; All of the possible options from $\tilde{Z}(\ell_{A+1})$ and $\tilde{Z}^ {ij}_2\of{x,\ell}$ are not shown.}
    \label{fig:plaq_uchain}
\end{figure*}
The plaquette update chain and its move choice probabilities are visualized in figure~\ref{fig:plaq_uchain}. The first part of the chain consists of plaquette updates that make $\Delta k_j$ and $\Delta \chi^{\of{i}}_j \bmod{2}$ zero to avoid the formation of defects when the temporal boundary conditions are changed. This is followed by an acceptance test for the actual change of the boundary conditions, taking into account the non-defect-causing changes the update would inflict on the site weights, due to non-zero $\Delta l_j=l_{\sigma\ssof{j}}-l_{j}$ and remnant (even) $\Delta \chi^{\of{i}}_j$ (cf. section~\ref{ssec:plaquetteupdate} for notation). The last part of the chain then consists of the same plaquette updates as the first part, but in reverse order, to restore $\Delta k_j$ and $\Delta \chi^{\of{i}}_j \bmod{2}$ to their original values, using the updated boundary conditions. For most of the chain, the selection probabilities are symmetric, since the probabilities for choosing a forward or a backward move are the same. This is, however, not the case when entering or exiting the entire update chain nor when entering or exiting a single plaquette update. 

At the start of each sweep, the program performs a random test of whether it should attempt a generalized worm or a boundary update. The boundary update is with a pre-set probability $p_b$. For our simulations, we set $p_b=0.25$. Once in the boundary update routine, the program will choose with equal probabilities whether to attempt to add or remove a site from region $A$. The total selection probability for an update chain that would change the size of $A$ in a given direction is therefore $p_b/2$. The selection probability for the move to exit the boundary update routine is set to be $1/2$, so that detailed balance requires this exit move to be accepted with probability $p_b$.

The fact that in each sweep, a boundary update can be attempted in addition to a generalized worm update also impacts the probabilities of the latter. The generalized worm update is now selected with probability $(1-p_b)$. As the type and insertion location of the source and sink pair is chosen randomly among the $N$ sectors and the $V_\Lambda$ lattice sites, the total selection probability of a particular source-sink insertion is $\frac{1-p_b}{N V_\Lambda}$. If the worm head and tail are at the same site, an attempt to remove them and terminate the worm update is selected with probability $1/2$. Then, denoting the configuration before the insertion as $C_0$ and the configuration after as $C_1$, the acceptance probabilities have to be 
\begin{equation} \label{eq:start_prob_general_worm}
    P_A(C_0\to C_1)=\min\left(1,\, \frac{NV_\Lambda}{2(1-p_b)}\frac{w(C_1)}{ w(C_0)}\right)
\end{equation}
and 
\begin{equation} \label{eq:stop_prob_general_worm}
    P_A(C_1\to C_0)=\min\left(1,\, \frac{2(1-p_b)}{NV_\Lambda}\frac{w(C_0)}{ w(C_1)}\right) \ ,
\end{equation}
for detailed balance to be satisfied. Note that the factor of $V_\Lambda$ in $P_A(C_0\to C_1)$ gets canceled by a relative factor of $1/V_{\Lambda}$ in $w(C_1)$ compared to $w(C_0)$, which is due to the factor of $1/V_{\Lambda}$ that multiplies the $Z_2^{i j}\of{x,y}$ terms in the definition of $Z_{\text{gen}}$ in \eqref{eq:zgen}. This definition of $Z_{\text{gen}}$, which removes the dependency on $V_{\Lambda}$ from \eqref{eq:start_prob_general_worm} and \eqref{eq:stop_prob_general_worm}, avoids low acceptance rates of the worm update termination move in large volumes and if the simulation is performed without optimized $c_{i\,j}$ weights.

As stated in section \ref{sect:latt_EE}, the spatial direction for a plaquette update has to be chosen such that the same temporal boundary conditions apply over the two spatial sites over which the temporal plaquette for the update is located. As long as this condition is satisfied, the direction can be freely chosen. For each plaquette, the spatial direction is therefore randomly chosen amongst the suitable options before the insertion of the head and tail pair is attempted. If the number of possible spatial directions is $n_p$, the move-choice probability for starting a specific plaquette is $\frac{1}{2n_p}$. The move-choice probability for terminating a plaquette worm is $1/2$ as in the case of the generalized worm. Putting these into \eqref{eq:MH_acc_prob}, we can see that the acceptance probabilities for starting the plaquette worm and terminating it have to be
\begin{equation}
    P_A(C_0\to C_1)=\min\left(1,\, n_p\frac{w(C_1)}{ w(C_0)}\right)
\end{equation}
and 
\begin{equation}
    P_A(C_1\to C_0)=\min\left(1,\, \frac{1}{n_p}\frac{w(C_0)}{ w(C_1)}\right)\ ,
\end{equation}
respectively. Here $C_1$ was used to denote the configuration after the insertion and $C_0$ the one before it. 

The configuration weights $w(C)$ have also been slightly modified in the plaquette updates. Firstly, the Wang-Landau biases $c_{ij}\of{x,y}$ from \eqref{eq:zgen} are not used. They have been optimized for the generalized worm to cover the entire system and as such can cause low acceptance rates when the worm is constrained to move along a plaquette. Secondly, the coupling of the temporal $k$-variable to the chemical potential in the partition function \eqref{eq:onfluxreppartf} is ignored. This can be done, since in a temporal plaquette, the worm always goes through exactly one link in both the positive and negative temporal direction. Therefore, the corresponding factors $\e^{\mu}$ and $\e^{-\mu}$ precisely cancel in the overall probability of the update. Ignoring the coupling to $\mu$ reduces fluctuations in the results and provides a cleaner signal.

\subsection{Boundary worm}

\begin{figure*}[!htb]
    \centering
    \includegraphics[scale=1.25]{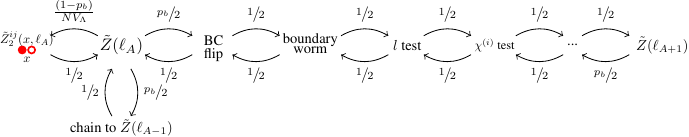}
    \caption{A visualization of the boundary worm update chain corresponding to adding a site to region $A$. In the actual program, the boundary worms and acceptance tests are placed to the chain in random order. The one here is just an example. The exact placement of the boundary flip is also somewhat arbitrary, since the worms are forbidden from going through the links affected by it. $\ell_{A+1}$ and $\ell_{A-1}$ are the $\ell$ values corresponding to adding and removing a site from $A$ respectively. The chains have been truncated at the end; All of the possible options from $\tilde{Z}(\ell_{A+1})$ and $\tilde{Z}^ {ij}_2\of{x,\ell}$ are not shown.}
    \label{fig:worm_uchain}
\end{figure*}
The update chain and movement choice probabilities of the boundary worm update are displayed in figure~\ref{fig:worm_uchain}. The chain consists of boundary worm updates that remove defects associated to non-zero $\Delta k_{j}$ and $\Delta\chi_{j}^{\of{i}}\bmod{2}$ caused by the change of temporal boundary conditions (cf. notation from section~\ref{ssec:plaquetteupdate}), and of acceptance tests for the $\Delta l_{j}$ and even parts of $\Delta\chi_j^{\of{i}}$, all in random order. The exact placement of the change of boundary conditions is arbitrary, since the worms are forbidden from going through the temporal links that change one of their endpoints under the change of boundary conditions. As with the plaquette, the movement choice probabilities are symmetric for the most part. There are two exceptions; when entering and exiting the update chain and when the worm head is moving in and out of the replica corners. 

For the boundary worm algorithm, the selection probabilities for moving in and out of the update chain are the same as for the plaquette boundary update. Therefore, the move to exit the update routine is once again accepted with probability $p_b$. 

A boundary worm can be terminated when the head is at either replica corner $x_j$, $j=0,\ldots,r-1$. When this is the case, an attempt to terminate the boundary worm is selected with probability $1/2$. Conversely, a specific head move out of this corner $x_j$ to some possible neighboring site $x$ is chosen with probability $1/\of{2u}$, where $u$ is the number of possible head moves, which is $u=2\,d$ or $u=2\,d+1$ if disconnected head moves are possible. Note that $u$ includes the forbidden head move along the link in negative time direction from $x_j$; if this move is selected, the acceptance probability is set to zero. The reverse move from $x$ to $x_j$ is selected with probability $1/u$. Using \eqref{eq:MH_acc_prob}, the acceptance probabilities for these moves must therefore be
\begin{equation}
    P_A(C_{x_j}\to C_{x})=\min\left(1,\, 2\frac{w(C_x)}{ w(C_{x_j})}\right)
\end{equation}
and 
\begin{equation}
    P_A(C_x\to C_{x_j})=\min\left(1,\, \frac{1}{2}\frac{w(C_{x_j})}{ w(C_x)}\right) \ ,
\end{equation}
respectively. Here, $C_{x_j}$ corresponds to the configuration with the worm head at $x_j$ and $C_{x}$ to the one with the head at $x$.

\section{Boundary update comparisons}\label{app:updates}

\begin{figure}[!htb]
    \centering
    \includegraphics[width=0.95\linewidth]{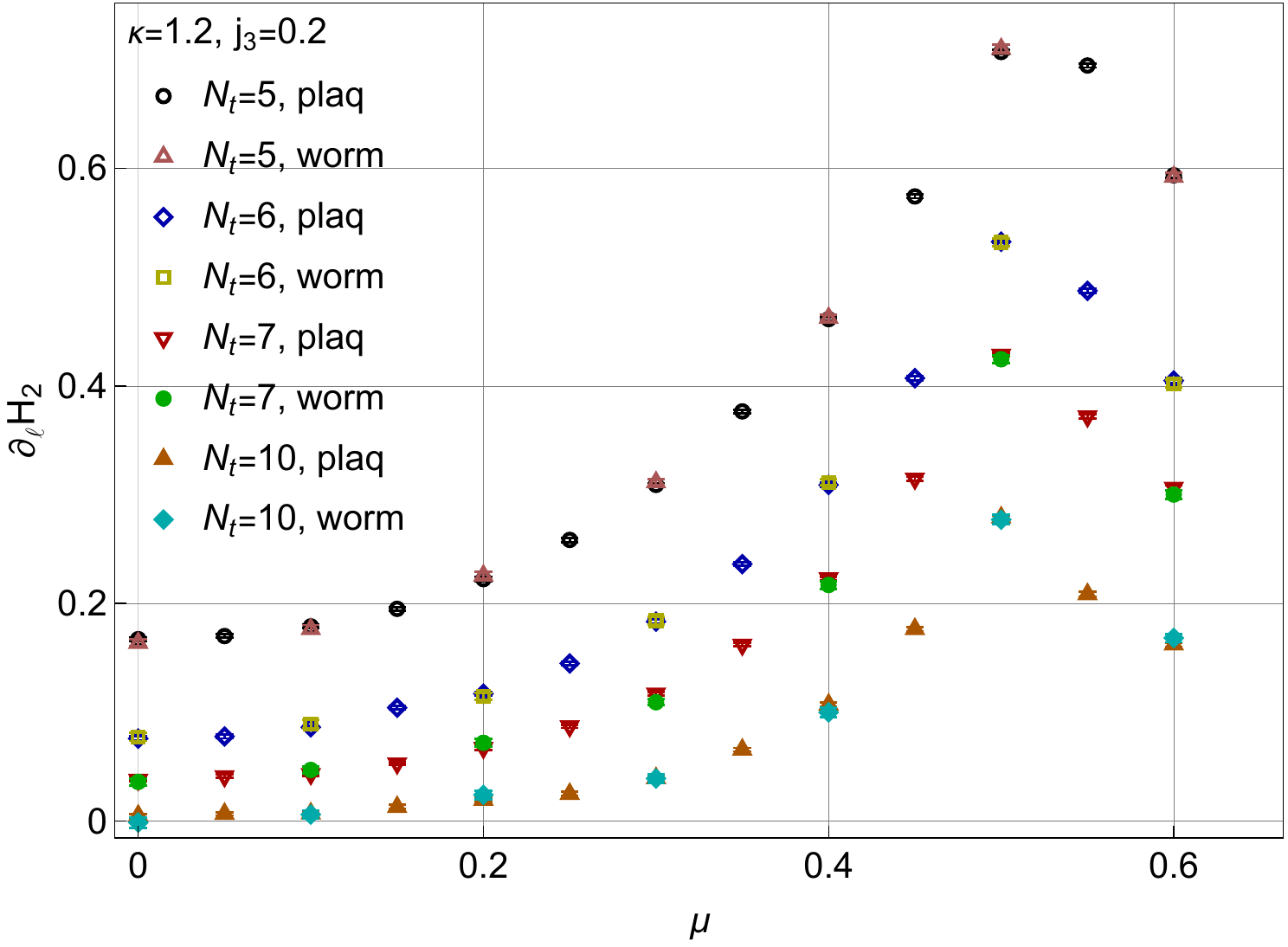}
    \caption{$\dHn{2}$ as a function of $\mu$ obtained with both the plaquette and the boundary worm updates. The results for the two updates show strong agreement.}
    \label{fig:plaq_worm_comparison}
\end{figure}

A comparison between results of $\dHn{2}$ as a function of $\mu$ at $N_t\in\{5,6,7,10\}$ obtained with the plaquette and boundary worm updates can be seen in figure~\ref{fig:plaq_worm_comparison}. The lattice sizes and parameters are the same as those described in section \ref{sec:select}. The results of the two updates match within the error bars, implying that both are working correctly.

To investigate the efficiency of both boundary update algorithms, we have measured their acceptance rates and update times with the simulation parameters used in this work. The acceptance rates are shown in table~\ref{tab:acc_rates}. We see that the acceptance rates for both updates are of similar magnitude, with the boundary worm having a slight edge. It is also observed that for both updates, the acceptance rate decreases with increasing $\mu$, while a dependency on $N_t$ seems to be absent.

\begin{table}[!htb]
\centering
\vspace{2mm}
 \begin{tabular}{|C{1.25cm}|C{1.25cm}|C{1.25cm}|C{2.0cm}|C{2.0cm}|} 
    \hline
    $N_t$&$\mu$&$T/m_0$&Acceptance rate for plaq&Acceptance rate for worm \\
    \hline
    5 & 0.3 &0.408& 0.273& 0.325 \\
    5 & 0.5 &0.408& 0.256 & 0.315 \\
     5 & 0.6&0.408& 0.198 & 0.296 \\
    10 & 0.3&0.204& 0.278 & 0.328 \\
     10 & 0.5&0.204& 0.261 & 0.319 \\
    10 &  0.6&0.204& 0.196 & 0.295 \\
    \hline
 \end{tabular}
 \caption{Acceptance rates for the plaquette, boundary worm and generalized worm updates at a few $N_t$ and $\mu$ values. The rates for the worm are a bit higher, but both are of similar magnitude. We see that the acceptance rate goes down with increasing $\mu$ for both updates.}
 \label{tab:acc_rates}
\end{table}

The update times are given in table ~\ref{tab:utimes} for both boundary updates and the generalized worm update. The exact values are hardware dependent, but their relative behavior should be fairly general. The shown values were obtained on Intel Xeon Gold 6230 CPUs of the Puhti supercomputer \cite{Puhti} operated by CSC - IT Center for Science, Finland. Of the boundary updates, the plaquette is faster across the board with little dependence on $N_t$. The update times for the boundary worm increase significantly with $N_t$. Therefore, we postulate that the difference between the update times will only increase with larger lattices. The update times for the generalized worm are similar to those of the boundary worm but slightly larger. They also seem to have a stronger $N_t$ dependence. This can be understood by recalling that the generalized worm samples a larger configuration space, which includes the off-diagonal two-point partition function channels requiring head-changing disconnected worm moves.

\begin{table}[!htb]
\centering
\vspace{2mm}
 \begin{tabular}{|C{0.55cm}|C{0.55cm}|C{2.35cm}|C{2.15cm}|C{2.25cm}|} 
    \hline
    $N_t$&$\mu$&Update time plaq [ms]&Update time boundary worm [ms]&Update time generalized worm [ms]\\
    \hline
    5 & 0.3 & $0.0300\pm0.0045$ & $1.77\pm0.016$ & $2.58\pm0.021$ \\
    5 & 0.5 & $0.0241\pm0.0034$ & $1.85\pm0.017$ & $2.61\pm0.020$\\
     5 & 0.6& $0.0386\pm0.0210$ & $2.09\pm0.019$ & $2.69\pm0.014$\\
    10 & 0.3& $0.0249\pm0.0047$ & $4.19\pm0.042$ & $6.18\pm0.052$\\
     10 & 0.5& $0.0159\pm0.0012$ & $4.08\pm0.043$ & $6.02\pm0.067$ \\
    10 &  0.6& $0.0173\pm0.0013$ & $4.83\pm0.053$ & $6.02\pm0.066$\\
    \hline
 \end{tabular}
 \caption{Update times for both boundary updates and the generalized worm update at a few $N_t$ and $\mu$ values. With the boundary updates, the update times for the plaquette are significantly smaller and seem to have little dependence on $N_t$. For the boundary worm, the update time approximately doubles for the doubled $N_t$. The generalized worm has update times similar to the boundary worm but slightly larger and with a stronger $N_t$ dependence, which can be understood by recalling that it samples a larger configuration space than the boundary worm.}
 \label{tab:utimes}
\end{table}

Of the two updates, the plaquette seems at first to be more efficient, as its update times are significantly shorter with only slightly lower acceptance rates. However, the plaquette updates are very localized and only ergodic in combination with the generalized worm updates. The boundary worm updates on the other hand update also the configuration variables in the bulk of the system in essentially the same way as the generalized worm update, even if the boundary condition update fails. However, no matter what boundary update algorithm is chosen, it seems reasonable to assume that the bulk should be updated more frequently than the boundary in order to minimize auto-correlation times. In our simulations the selection probability ratio between generalized worm and boundary updates was chosen to be 3:1, which resulted in the program spending 80\% or more of the simulation time in the generalized worm updates. The choice of boundary update algorithm was therefore of secondary importance when considering the overall performance and efficiency of the simulation algorithm. Future efforts to optimize our simulation program will primarily focus on parallelizing the bulk update, \ie, the generalized worm.  

While in this work, each simulation relied merely on either the plaquette or the boundary worm update to update the boundary conditions, the two methods could in principle also be mixed. Each individual boundary update would still need to use only one type at a time, but one could randomly alternate (with tunable probabilities) between the two types in order to optimize the efficiency of the simulation, \ie, to reduce auto-correlation times. We will leave this for future studies.  



\bibliography{refs}

\end{document}